\documentclass[draftclsnofoot, journal, letterpaper, onecolumn, 12pt]{IEEEtran}
\usepackage{amsmath,graphicx}
\usepackage{algorithm}
\usepackage{algpseudocode}
\usepackage{amssymb}
\usepackage{subfigure}
\usepackage{footmisc}
\usepackage{lipsum}
\usepackage{mathtools}
\usepackage{cuted}
\usepackage{esint}
\usepackage{multicol}
\usepackage{multirow}
\usepackage{cite}
\usepackage{xcolor}
\usepackage{accents}
\usepackage{etoolbox}

\newcommand\munderbar[1]{%
  \underaccent{\bar}{#1}}

\makeatletter
\@for\next:={int,iint,iiint,iiiint,dotsint,oint,oiint,sqint,sqiint,
  ointctrclockwise,ointclockwise,varointclockwise,varointctrclockwise,
 fint,varoiint,landupint,landdownint}\do{%
    \expandafter\edef\csname\next\endcsname{%wjdalsc
   \noexpand\DOTSI
      \expandafter\noexpand\csname\next op\endcsname
   \noexpand\ilimits@
    }%
 }
\makeatother
\begin{document}

\title{On the Optimality of Reconfigurable Intelligent Surfaces (RISs): Passive Beamforming, Modulation, and Resource Allocation}

\author{Minchae~Jung,~\IEEEmembership{Member,~IEEE}, Walid~Saad,~\IEEEmembership{Fellow,~IEEE}, \\
Mérouane~Debbah,~\IEEEmembership{Fellow,~IEEE},
and Choong~Seon~Hong,~\IEEEmembership{Senior Member,~IEEE}% <-this % stops a space

\thanks{
{\hspace{-0.46cm}M. Jung is with the Department of Electronic Engineering, Soonchunhyang University, Asan, Chungcheongnam-do, Rep. of Korea (e-mail: hosaly@sch.ac.kr).}

{W. Saad is with Wireless@VT, Department of Electrical and Computer Engineering, Virginia Tech, Blacksburg, VA 24061 USA (e-mail: walids@vt.edu).}

{M. Debbah is with CentraleSupelec, Université Paris-Saclay, 91192 Gifsur-Yvette, France and Mathematical \& Algorithmic Sciences Lab, Huawei Technologies France SASU, 92100 Boulogne-Billancourt, France (e-mail: merouane.debbah@huawei.com).}

{C. Hong is with the Department of Computer Science and Engineering, Kyung Hee University, Yongin-si, Gyeonggi-do 17104, Rep. of Korea (e-mail: cshong@khu.ac.kr).}
\vspace{-0.05cm}
%This research was supported by Basic Science Research Program through the National Research Foundation of Korea (NRF) funded by the %Ministry of Education (NRF-2016R1A6A3A11936259).
%This research was supported by Basic Science Research Program through the National Research Foundation of Korea (NRF) funded by the Ministry of Education (NRF-2016R1A6A3A11936259) and by the U.S. National Science Foundation under Grants CNS-1836802 and OAC-1638283.
}
}

% The paper headers
\markboth{}%'
{Shell \MakeLowercase{\textit{et al.}}: Bare Demo of IEEEtran.cls for Journals}

\maketitle

\vspace{-1.85cm}
\begin{abstract}
Reconfigurable intelligent surfaces (RISs) have recently emerged as a promising technology that can achieve high spectrum and energy efficiency for future wireless networks by integrating a massive number of low-cost and passive reflecting elements.
An RIS can manipulate the properties of an incident wave, such as the frequency, amplitude, and phase,
and, then, reflect this manipulated wave to a desired destination, without the need for complex signal processing.
In this paper, the asymptotic optimality of achievable rate in a downlink RIS system is analyzed under a practical RIS environment with its associated limitations.
In particular, a passive beamformer that can achieve the asymptotic optimal performance by controlling the incident wave properties is designed, under a limited RIS control link and practical reflection coefficients.
In order to increase the achievable system sum-rate, a modulation scheme that can be used in an RIS without interfering with existing users is proposed
and its average symbol error rate is asymptotically derived.
Moreover, a new resource allocation algorithm that jointly considers user scheduling and power control is designed,
under consideration of the proposed passive beamforming and modulation schemes.
Simulation results show that the proposed schemes are in close agreement with their upper bounds in presence of a large number of RIS reflecting elements
thereby verifying that the achievable rate in practical RISs satisfies the asymptotic optimality.
%and even can achieve the performance higher than existing upper bound resulting from the additional data rates at unscheduled users.
\end{abstract}

\vspace{-0.4cm}
\begin{IEEEkeywords}
\vspace{-0.5cm}
Reconfigurable intelligent surface (RIS), metasurface, passive beamforming, resource allocation, 

\end{IEEEkeywords}

\vspace{-0.5cm}
\IEEEpeerreviewmaketitle
\section{Introduction}\vspace{-0.1cm}
\IEEEPARstart{T}{he} concept of a metasurface is rapidly emerging as a key solution to support the demand for massive connectivity, mainly driven by upcoming Internet of Things (IoT) and 6G applications \cite{ref.Wu2019bfoptimization,ref.Saad2019vision,ref.Mozaffari2019beyond,ref.Jung2019reliability,ref.Jung2018lisul,ref.Chaccour2020risk,ref.Cui2017information,ref.Basar2019wireless,
ref.Tang2019program,ref.Tang2018wireless,ref.Wu2018beamforming,ref.Han2018assisted,ref.Huang2018energy,ref.Jung2019tbsICC,ref.Siev2003two,ref.Zhu2012active,
ref.Hu2018data,ref.R1A2,ref.R1A4}.
A metasurface relies on a massive integration of artificial meta-atoms that are commonly made of metal structures of low-cost and passive elements \cite{ref.R2A3, ref.R2A4}.
Each meta-atom can manipulate the incident electromagnetic (EM) wave impinging on it, in terms of frequency, amplitude, and phase,
and reflect it to a desired destination, without additional signal processing.
A metasurface can potentially provide reliable and pervasive wireless connectivity given that man-made structures, such as buildings, walls, and roads, can be equipped with metasurfaces in the near future
and used for wireless transmission \cite{ref.Jung2019reliability,ref.Jung2018lisul,ref.Chaccour2020risk}.
Moreover, a tunable metasurface can significantly enhance the signal quality at a receiver
by allowing a dynamic manipulation of the incident EM wave.
%a metasurface needs to be controlled for varying wireless channels
%and a tunable metasurface can offer dynamic manipulates of the EM wave to coherently align with those channels.
%Moreover, the EM wave received at the desired destination can be enhanced by manipulating the incident EM wave to coherently align with the desired channel, under consideration of tunable metasurfaces.
Tunable metasurfaces are mainly controlled by electrical, optical, mechanical, and fluid operations \cite{ref.Cui2017information}
that can be programmed in software using a field programmable gate array (FPGA) \cite{ref.Basar2019wireless}.
The concept of a reconfigurable intelligent surface (RIS) is essentially an electronically operated metasurface controlled by programmable software, as introduced in \cite{ref.Cui2017information} and \cite{ref.Basar2019wireless}.
In wireless communication systems, a base station (BS) can send control signals to an RIS controller (i.e., FPGA) via a dedicated control link and controls the properties of the incident wave to enhance the signal quality at the receiver.
%More practically, in wireless communication systems, a base station (BS) sends control signals to the RIS controller (e.g., FPGA) via a dedicated control link
%and controls the properties of incident wave according to their purpose.
%In principle, the electrical size of the unit reflecting elements (i.e., meta-atoms) deployed on such RISs is between $\lambda/8$ and $\lambda/4$, where $\lambda$ is a wavelength of radio frequency (RF) signal \cite{ref.Cui2017information}.
%Note that conventional large antenna-array systems, such as massive multiple-input and multiple-output (MIMO) and MIMO relays, typically require antenna spacing of greater than $\lambda/2$ in order to minimize the coupling loss and obtain full diversity gain \cite{ref.Jung2018lisul}.
%Therefore, an RIS can provide more reliable and space-intensive communications compared to conventional antenna-array systems as clearly explained in \cite{ref.Jung2019reliability,ref.Jung2018lisul,ref.Jung2019spectral}.
%More practically, in wireless communication systems, a base station (BS) sends control signals to the RIS controller (e.g., FPGA) via a dedicated control link
%and controls the properties of incident wave according to their purpose.
In principle, the electrical size of the unit reflecting elements (i.e., meta-atoms) deployed on RIS is between $\lambda/8$ and $\lambda/4$, where $\lambda$ is a wavelength of radio frequency (RF) signal \cite{ref.Cui2017information}.
Note that conventional large antenna-array systems, such as a massive multiple-input and multiple-output (MIMO) and MIMO relay system, typically require antenna spacing of greater than $\lambda/2$ \cite{ref.Jung2018lisul}. %in order to minimize the coupling loss and obtain full diversity gain \cite{ref.Jung2018lisul}.
Therefore, an RIS can provide more reliable and space-intensive communications compared to conventional antenna-array systems as clearly explained in \cite{ref.Jung2019reliability,ref.Jung2018lisul,ref.Chaccour2020risk}.
Therefore, a large number of reflecting elements can be arranged on each RIS thus offering precise control of the reflection wave and allowing it to coherently align with the desired channel.
Furthermore, an RIS can be used with other promising communication techniques, such as simultaneous wireless information and power transfer (SWIPT) \cite{ref.WPT00, ref.WPT01}
and physical layer security \cite{ref.PSS00}.
Hence, the use of RIS has been recognized as a promising technology for future 6G wireless systems \cite{ref.w1,ref.w2,ref.w3}.

\vspace{-0.45cm}
\subsection{Related Works}\vspace{-0.1cm}
Owing to these advantages, the use of an RIS in wireless communication systems has recently received significant attention as in \cite{ref.Wu2019bfoptimization} and \cite{ref.Basar2019wireless,ref.Wu2018beamforming,ref.Tang2019program,ref.Tang2018wireless,ref.Han2018assisted,ref.Huang2018energy,ref.Jung2019tbsICC}.
An RIS is typically used for two main wireless communication purposes: a) RIS as an RF chain-free transmitter and b) RIS as a passive beamformer that amplifies the incident waveform (received from a BS) and reflects it to the desired user, so-called as intelligent reflecting surface (IRS).
In \cite{ref.Basar2019wireless}, the authors analyzed the error rate performance of a phase-shift keying (PSK) signaling
and proved that an RIS transmitter equipped with a large number of reflecting elements can convey information with high reliability.
The works in \cite{ref.Tang2019program} and \cite{ref.Tang2018wireless} proposed RF chain-free transmitter architectures enabled by an RIS that can support PSK and quadrature amplitude modulation (QAM).
Meanwhile, the works in \cite{ref.Wu2019bfoptimization} and \cite{ref.Wu2018beamforming} designed joint active and passive beamformer that minimize the transmit power at the BS, under discrete and continuous phase shifts, respectively.
Also, in \cite{ref.Han2018assisted} and \cite{ref.Huang2018energy}, the authors designed a passive beamformer that maximizes the ergodic data rate and the energy efficiency, respectively. 
Moreover, the work in \cite{ref.Basar2019wireless} theoretically analyzed the average symbol error rate (SER) resulting from an ideal passive beamformer
and proved that the SER decays exponentially as the number of reflecting elements on RIS increases.
In \cite{ref.Jung2019tbsICC}, we provided a first insight on a passive beamformer that can achieve, asymptotically, an ideal RIS performance.
However, these previous studies in \cite{ref.Wu2019bfoptimization} and \cite{ref.Basar2019wireless,ref.Wu2018beamforming,ref.Jung2019tbsICC,ref.Tang2019program,ref.Tang2018wireless,ref.Han2018assisted,ref.Huang2018energy} \emph{have not considered practical RIS environments and their limitations, such as practical reflection coefficients and the limited capacity of the RIS control link.}
In fact, an RIS can manipulate the properties of an incident wave based on the resonant frequency of the tunable reflecting circuit.
Then, the incident EM power is partially consumed at the resistance of the reflecting circuit according to the difference between the incident wave frequency and the resonant frequency.
This results in the amplitude of the reflection coefficients less than or equal to one depending on the phase shifts of the incident wave. %in an RIS.
However, the works in \cite{ref.Wu2019bfoptimization, ref.Basar2019wireless,ref.Wu2018beamforming} and \cite{ref.Han2018assisted} assumed an ideal RIS whose amplitude of the reflection coefficients are always equal to one which is impractical for an RIS.
Moreover, in \cite{ref.Basar2019wireless} and \cite{ref.Wu2018beamforming,ref.Jung2019tbsICC,ref.Han2018assisted,ref.Huang2018energy},
the authors assumed a continuous phase shift at each reflecting element.
However, this continuous phase shift requires infinite bits to control each reflecting element
and the RIS control link between a BS and an RIS cannot support those infinite control bits.
Finally, the signals from the RIS transmitters proposed in \cite{ref.Tang2019program} and \cite{ref.Tang2018wireless} can be undesired interference for existing cellular network,
given that those RISs operate as an underlay coexistence with cellular networks.
Therefore, there is a need for new analysis of practical RISs when dealing with a limited RIS control link capacity and practical reflection coefficients
that can verify the asymptotic optimality of realistic RISs.

\vspace{-0.55cm}
\subsection{Contributions}\vspace{-0.1cm}
The main contribution of this paper is a rigorous optimality analysis of the data rates that can be achieved by an RIS under consideration of practical reflection coefficients with a limited RIS control link capacity\footnote{An asymptotic optimal performance can be achieved by an ideal RIS which has infinite and continuous phase shifts and lossless reflection coefficients
and is higher than that of a massive MIMO system as proved in \cite{ref.Wu2019bfoptimization}.}.
In this regard, we first design a passive beamformer that achieves asymptotic signal-to-noise ratio (SNR) optimality, regardless of the reflection power loss and the number of RIS control bits.
In particular, the proposed passive beamformer with one bit RIS control can achieve the asymptotic SNR of an ideal RIS with infinite control bits, lead to a much simpler operation at the BS especially for a large number of reflecting elements on an RIS.
We then propose a new modulation scheme that can be used in an RIS to achieve sum-rate higher than the one achieved by a conventional network without RIS. %, without using additional radio resource.
In the proposed modulation scheme, each RIS utilizes an ambient RF signal, convert it into desired signal by controlling the properties of incident wave, and transmit it to the desired user, without interfering with existing users.
We also prove that the achievable SNR from the proposed modulation converges to the asymptotic SNR resulting from a conventional massive MIMO or MIMO relay system, as the number of reflecting elements on an RIS increases.
Given the aforementioned passive beamformer and modulation scheme, we finally develop a novel resource allocation algorithm whose goal is to maximize the average sum-rate under the minimum rate requirements at each user.
We then study, analytically, the potential of an RIS by showing that a practical RIS can achieve the asymptotic performance of an ideal RIS, as the number of RIS reflecting elements increases without bound.
Our simulations show that the proposed schemes can asymptotically achieve the performance resulting from an ideal RIS and its upper bound.

The rest of this paper is organized as follows. 
Section II describes the system model.
Section III describes the optimality of achievable rate in downlink RIS system.
Simulation results are provided in Section IV to support and verify the analyses, and
Section V concludes the paper.

\textit{Notations:} Throughout this paper, boldface upper- and lower-case symbols represent matrices and vectors respectively, and $\boldsymbol{I}_M$ is a size-$M$ identity matrix.
$\left({\boldsymbol{A}} \right)_{i,j}$ refers to an element at row $i$ and column $j$ of a matrix ${\boldsymbol{A}}$
%${\mu _X}$ and $\sigma _X^2$ denote the mean and variance of a random variable $X$, respectively.
%\begin{figure}[!ht]
%\centering
%\includegraphics[width=0.61\columnwidth] {figs/fig_lis_structure.eps}\vspace{-0.3cm}
%\caption{Illustrative architecture of an RIS consist of multiple two-dimensional layers.}\vspace{-0.7cm}
%\label{fig.1}
%\end{figure}
The conjugate, transpose, and Hermitian transpose operators are ${\left( \cdot \right)^*}$, ${\left( \cdot \right)^{\rm{T}}}$, and ${\left( \cdot \right)^{\rm{H}}}$, respectively. 
The norm of a vector $\boldsymbol{a}$ is $\left\| {\boldsymbol{a}} \right\|$,
the amplitude and phase of a complex number $a$ are denoted by $\left| {{a}} \right|$ and $\angle{{a}}$, respectively.
${\rm{E}}\left[ \cdot \right]$, ${\rm{Var}}\left[ \cdot \right]$, and ${\rm{Cov}}\left[ \cdot \right]$ denote expectation, variance, and covariance operators, respectively.
%$\mathcal{O}\left( \cdot \right)$ and $ \otimes$ and denote the big O notation and the Kronecker product.
%The operators ${\mathop{\rm Re}\nolimits} \left( \cdot \right)$ take the real part.
$\mathcal{O}\hspace{-1pt}\left( \cdot \right)$ denotes the big O notation 
and $\mathcal{CN}\hspace{-2pt}\left( {m,{\sigma ^2}} \right)$ denotes a complex Gaussian distribution with mean $m$ and variance $\sigma ^2$.

\begin{figure}[t]
\centering
\includegraphics[width=0.53\columnwidth] {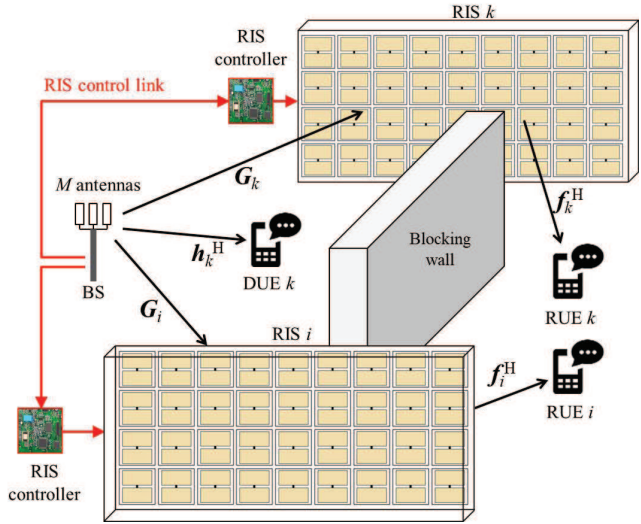}\vspace{-0.4cm}
\caption{Illustrative system model of the considered RIS-based MISO system.}\vspace{-0.7cm}
\label{fig.7}
\end{figure}

\vspace{-0.3cm}
\section{System Model}
Consider a single BS multiple-input single-output (MISO) system that consists of a set $\mathcal{K}$ of $K$ single-antenna user equipments (UEs) and multiple RISs each of which having $N$ reflecting elements, as shown in Fig. {\ref{fig.7}}.
The BS is equipped with $M$ antennas and serves one UE at each resource block (RB) based on an orthogonal frequency-division multiple-access (OFDMA) scheme.
We assume that the total system bandwidth is equally divided into $C$ orthogonal subcarriers and $F$ orthogonal RBs, indexed by $c \in \mathcal{C}= {\left[ {0, \cdots ,{C-1}} \right]}$ and $f \in   \mathcal{F}={\left[ {0,\cdots ,{F-1}} \right]}$, respectively.
Also, the BS transmits the downlink signal to scheduled UE through the transmit beamforming.
%Although we consider a TDMA system, we can readily extend the proposed model into an orthogonal frequency-division multiple-access (OFDMA) system. %%%%%%%%%%%%%%%%%%%%%%%%%%%%%%%%%%%%%%%%%%%%
In our system model, we consider two types of UEs: a) UEs directly connected to the BS (called DUEs) and b) UEs connected to the BS via an RIS (called RUEs).
Each UE can measure the downlink channel quality information (CQI) and transmit this information to the BS as done in existing cellular systems \cite{ref.LTE2017TS36521}.
For UEs whose CQI exceeds a pre-determined threshold, the BS will directly transmit downlink signals to these UEs (which are now DUEs) without using the RIS.
When the CQI is below a pre-determined threshold (i.e., the direct BS-UE channel is poor), the BS will have to allocate, respectively, suitable RISs to those UEs (that become RUEs) experiencing this poor CQI and, then, send a control signal to each RIS controller via a dedicated control link.
Given the received BS control signal, the RIS controller determines $N$ bias direct-current (DC) voltages for all reflecting elements and then, the varactor capacitance can be controlled, resulting in phase shifts of the reflection wave.
We ignore the signals that are reflected by the RIS more than once because their signal power
will be negligible due to severe path loss and the reflection power loss.
Note that an RIS cannot coherently align, simultaneously, with the desired channels of all RUEs which, in turn, limits system performance \cite{ref.Jung2019reliability,ref.Jung2018lisul,ref.Chaccour2020risk}.
For densely located RISs,
we assume that each RUE is connected to different RISs depending on the location of each RUE. %(i.e., one RUE per one RIS per one RB) 
%\textcolor{blue}{
%If one RUE is scheduled multiple times within a time slot with different RBs, that RUE will be connected to different RISs for different RBs.}
%For notational simplicity, we assume that RUE $k$ is connected to RIS $k$.
%Consider an RIS based multiple-input single-output (MISO) system that consists of an RIS having $N$ reflecting elements that reflect the downlink signal from a BS to a set $\mathcal{K}$ of $K$ user equipments (UEs), as shown in Fig. {\ref{fig.7}}.
%The AP equipped with $M$ antennas serves a single-antenna UE based on a time-division multiple access (TDMA) and transmits downlink signal to the UE through the transmit beamforming.
%Therefore, the direct path between the AP and RUE can be negligible because of unfavorable channel conditions.
\begin{figure}
\centering
\includegraphics[width=0.6\columnwidth] {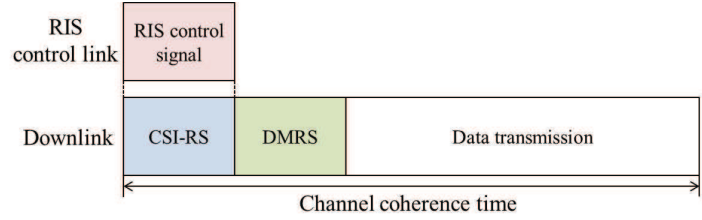}\vspace{-0.6cm}
\caption{Downlink frame structure at BS in considered RIS system.}\vspace{-0.7cm}
\label{fig.8}
\end{figure}
Also, given a practical range of mobility and carrier frequencies, 
we consider that all channels are generated from a quasi-static block fading model whose coherence bandwidth and time cover the bandwidth of an RB and its downlink transmission period.
Hence, the wireless channel and the SNR can be assumed approximately constant within each RB as done in  \cite{ref.Chiani2011OFDM}.
In accordance with 3GPP LTE specification \cite{ref.LTE2018TS38214}, we consider two types of reference signals (RSs): Channel state information-RS (CSI-RS) and demodulation RS (DMRS). 
We use a CSI-RS to estimate the CSI and measure the CQI, and we use a DMRS, which is a beamformed RS, in order to estimate an effective CSI for demodulation \cite{ref.LTE2018TS38214}.
In order to accurately estimate the CSI, the RIS does not shift, artificially, the phase of the incident wave during the CSI-RS period\footnote{Note that a perfect electric conductor (PEC) basically shifts the reflection phase of the incident wave by $\pi$ without reflection power loss \cite{ref.Zhu2012active}. We assume that the surface of each RIS element is made of a metallic PEC.}
%(which equals to $\pi$ phase shift of incident wave resulting from the wave reflection ifself \cite{ref.Zhu2012active}), 
and, simultaneously, the BS can send a control signal to the RIS controller via a dedicated control link during this period.
Then, the RIS operates based on this control signal and reflects the DMRS and data signal with controlled phase shifts. %(i.e., passive beamforming).
The RUE receives the DMRS and estimates the phase-shifted CSI, and eventually, the downlink signal can be decoded.
Hereinafter, we assume that the channel state of each wireless link follows a stationary stochastic process
under a perfect CSI at the BS 
and, hence, our analysis will result in a performance bound of practical channel estimation scenarios.

We divide the UE set into two sets, such that $\mathcal{K} = \mathcal{D} \cup \mathcal{R}$ where $\mathcal{D}$ is the set of DUEs and $\mathcal{R}$ is the set of RUEs.
Then, the received signal at UE $k$ is obtained as
\begin{equation}
{y_k} = \left\{ \begin{array}{l}
\sqrt{P}{\boldsymbol{h}_k^{\rm{H}}}{\boldsymbol{w}_k}{x_k^{\rm{d}}} + {n_k^{\rm{d}}},\hspace{1.08cm}{\text{ if }} k \in \mathcal{D} ,\\
\sqrt{P}\boldsymbol{f}_k^{\rm{H}}{\boldsymbol{\Phi} _k}\boldsymbol{G}_k{\boldsymbol{w}_k}{x_k^{\rm{r}}} + {n_k^{\rm{r}}}, {\text{ if }} k \in \mathcal{R},
\end{array} \right. \label{eq.yk}
\end{equation}
\hspace{-6pt} where $P$ is the BS transmit power and ${\boldsymbol{h}}_k \hspace{-1pt}\in\hspace{-1pt}\mathbb{C}^{M \times 1}$, $\boldsymbol{G}_k \hspace{-1pt}\in\hspace{-1pt} \mathbb{C}^{N \times M}$, and $\boldsymbol{f}_k\hspace{-1pt} \in\hspace{-1pt}\mathbb{C}^{N \times 1}$ are, respectively, the fading channels between the BS and DUE $k$, 
between the BS and RIS $k$, and between RIS $k$ and RUE $k$.
In fact, all channels, matrices, vectors, and signals presented in (\ref{eq.yk}) should include the corresponding OFDM symbol index and the subcarrier index.
However, for notational simplicity, we omit them here, but we will use them in Section IV. C.
For a practical RIS environment, $\boldsymbol{G}_k$ and $\boldsymbol{f}_k$ can be generated by Rician fading composed of a deterministic line-of-sight (LoS) path and spatially correlated non-LoS (NLoS) path:
\begin{align}
%{{\boldsymbol{h}}_{k}} &= \sqrt {\frac{{{\kappa _{{\rm{B}}k}}}}{{{\kappa _{{\rm{B}}k}} + 1}}} {\boldsymbol{h}}_{{\rm{B}}k}^{\rm{L}} + \sqrt {\frac{1}{{{\kappa _{{\rm{B}}k}} + 1}}}{\boldsymbol{R}}^{1/2}_{{\rm{B}}k} {\boldsymbol{h}}_{{\rm{B}}k}^{{\rm{NL}}},\\
{{\boldsymbol{G}}_{k}} &= \sqrt {\frac{{{\kappa _{{\rm{b}}k}}}}{{{\kappa_{{\rm{b}}k}} + 1}}} {\boldsymbol{\bar G}}_{k} + \sqrt {\frac{1}{{{\kappa _{{\rm{b}}k}} + 1}}}{\boldsymbol{R}}_{{\rm{b}}k}^{1/2} {\boldsymbol{\tilde G}}_{k},\label{eq.Gk}\\
{{\boldsymbol{f}}_{k}} &= \sqrt {\frac{{{\kappa _{{\rm{r}}k}}}}{{{\kappa _{{\rm{r}}k}} + 1}}} {\boldsymbol{\bar f}}_{k}+ \sqrt {\frac{1}{{{\kappa _{{\rm{r}}k}} + 1}}} {\boldsymbol{R}}_{{\rm{r}}k}^{1/2} {\boldsymbol{\tilde f}}_{k},\label{eq.fk}
\end{align}
where $\kappa _{{\rm{b}}k}$ and $\kappa _{{\rm{r}}k}$ are the Rician factors between the BS and RIS $k$ and between RIS $k$ and RUE $k$, respectively,
and ${\boldsymbol{R}}_{{\rm{b}}k} \hspace{-1pt}\in\hspace{-1pt} \mathbb{C}^{N \times N}$ and ${\boldsymbol{R}}_{{\rm{r}}k} \hspace{-1pt}\in\hspace{-1pt} \mathbb{C}^{N \times N}$ are their spatial correlation matrices.
Also, ${\boldsymbol{\bar G}}_{k} \hspace{-1pt}\in\hspace{-1pt} \mathbb{C}^{N \times M}$ and ${\boldsymbol{\bar f}}_{k} \hspace{-1pt}\in\hspace{-1pt} \mathbb{C}^{N \times 1}$ are the deterministic LoS components
and ${\boldsymbol{\tilde G}}_{k} \hspace{-1pt}\in\hspace{-1pt} \mathbb{C}^{N \times M}$ and ${\boldsymbol{\tilde f}}_{k} \hspace{-1pt}\in\hspace{-1pt} \mathbb{C}^{N \times 1}$ are the NLoS components
whose entries are i.i.d. complex Gaussian random variables with zero mean and unit variance.
We also assume that ${\boldsymbol{h}}_k$ is generated by uncorrelated Rician fading composed of a deterministic LoS path and spatially uncorrelated NLoS path under the assumption that the BS antenna spacing is considerably larger than the carrier wavelength  in a relatively non-scattering environment, as follows:
\begin{equation}
{{\boldsymbol{h}}_{k}} = \sqrt {\frac{{{\kappa _{{\rm{d}}k}}}}{{{\kappa _{{\rm{d}}k}} + 1}}} {\boldsymbol{\bar h}}_{k}+ \sqrt {\frac{1}{{{\kappa _{{\rm{d}}k}} + 1}}}{\boldsymbol{\tilde h}}_{k},\label{eq.hk}
\end{equation}
where $\kappa _{{\rm{d}}k}$ is the Rician factor between the BS and DUE $k$, ${\boldsymbol{\bar h}}_{k} \hspace{-1pt}\in\hspace{-1pt} \mathbb{C}^{M \times 1}$ is the deterministic LoS component, and ${\boldsymbol{\tilde h}}_{k} \hspace{-1pt}\in\hspace{-1pt} \mathbb{C}^{M \times 1}$ is the NLoS component.
${\boldsymbol{w}_k}\in \mathbb{C}^{M \times 1}$ is the transmit beamforming vector that changes both the phase and amplitude of the effective downlink channel,
 and $x_k^{\rm{d}}$ and $x_k^{\rm{r}}$ are downlink transmit symbols for DUE $k$ and RUE $k$, respectively, with noise terms $n_k^{\rm{d}} \sim \mathcal{CN}\left(0,N_0\right)$ and $n_k^{\rm{r}} \sim \mathcal{CN}\left(0,N_0\right)$.
In (\ref{eq.yk}), ${\boldsymbol{\Phi} _k}\in \mathbb{C}^{N \times N}$ is a reflection matrix (i.e., passive beamformer) that includes reflection amplitudes and phases resulting from $N$ reflecting elements.
This reflection matrix is controlled by the RIS control signal from the BS and then, ${\boldsymbol{\Phi} _k}$ can be obtained as follows:
\begin{equation}
{\boldsymbol{\Phi} _k}={\rm{diag}}\left( { A\left(\angle \Gamma_1^k\right)e^{j\angle {\Gamma _1^k}}, A\left(\angle \Gamma_2^k\right)e^{j\angle {\Gamma _2^k}}, \cdots , A\left(\angle \Gamma_N^k\right)e^{j\angle {\Gamma _N^k}}} \right),\label{eq.PHIk}
\end{equation}
where $A\left(\angle \Gamma_n^k\right)$ and $\angle {\Gamma _n^k}$ are the reflection amplitude and phase at reflecting element $n$ of RIS $k$, respectively.
We consider a practical reflection power loss resulting from the power consumption at the resistance of the reflecting circuit.
In \cite{ref.Abey2019intelligent}, the relation between a reflection amplitude and its phase is approximated under this practical reflection power loss, as follows:%by using a standard curve fitting tool, as follows: 
\begin{equation}
A\left(\angle \Gamma_n^k\right)=\left(1- \left| {{\Gamma}} \right|_{\rm{min}}\right){\left( {\frac{{\sin \left( {\angle {\Gamma _n^k} - 0.43\pi } \right) + 1}}{2}} \right)^{1.6}} + \left| {{\Gamma}} \right|_{\rm{min}}, \label{eq.approx}
\end{equation}
where $\left| {{\Gamma}} \right|_{\rm{min}}=0.2$ is the minimum reflection amplitude when we use the SMV1231-079 varactors in the RIS \cite{ref.Abey2019intelligent}.
Hereinafter, we use the this reflection model and will verify  asymptotic optimality of practical RISs.
Hence, the instantaneous SNR at UE $k$ is given as follows:
\begin{equation}
{\gamma_k} = \left\{ \begin{array}{l}
{PE_{{k}}^{\rm{d}}}\left|{\boldsymbol{h}_k^{\rm{H}}}{\boldsymbol{w}_k}\right|^2 / N_0,\hspace{1.055cm}{\text{ if }} k \in \mathcal{D} ,\\
{PE_{{k}}^{\rm{r}}}\left|\boldsymbol{f}_k^{\rm{H}}{\boldsymbol{\Phi} _k}\boldsymbol{G}_k {\boldsymbol{w}_k}\right|^2/ N_0,  {\text{ if }} k \in \mathcal{R} ,
\end{array} \right. \label{eq.SNRk}
\end{equation}
where $E_k^{\rm{d}}$ and $E_k^{\rm{r}}$ are the average energy per symbol for DUE $k$ and RUE $k$, respectively.
Given this practical RIS model, our goal is to maximize (\ref{eq.SNRk}) and eventually achieve (asymptotically) the SNR of an ideal RIS as $N \to \infty$.
In most prior studies such as \cite{ref.Wu2019bfoptimization} and \cite{ref.Wu2018beamforming,ref.Han2018assisted,ref.Huang2018energy},
\emph{the properties of the reflection wave, such as the frequency, amplitude, and phase, are assumed to be independently controlled,
however, these properties are closely related to each other as discussed in Section II. A.}
Hence, their relationship should be considered in the system model to accurately verify the potential of practical RISs.
Note that the SNR of an ideal RIS system increases with $\mathcal{O}\left(N^2 \right)$ as $N$ increases \cite{ref.Wu2018beamforming}.
Since the diversity order of a conventional antenna array system is linearly proportional to the number of transmit antennas \cite{ref.Ngo2013energy},
an ideal RIS can achieve a squared diversity order of a conventional array system equipped with $N$ transmit antennas.
This squared diversity gain can be obtained from an ideal RIS assumption in which the incident wave is reflected by an RIS without power loss
and the BS sends infinite bits to the RIS controller via unlimited RIS control link capacity.
Given a practical RIS model in (\ref{eq.SNRk}), we will propose a novel passive beamformer that can achieve $\mathcal{O}\left(N^2 \right)$, asymptotically, even with one bit control for each reflecting element.
Moreover, we will propose a new modulation scheme and an effective resource allocation algorithm that can be used in our RIS system to increase an achievable sum-rate under the aforementioned practical RIS considerations.

\vspace{-0.5cm}
\section{Optimality of the Achievable Downlink Rate in an RIS}\vspace{-0.2cm}
We analyze the optimality of the achievable rate using practical RISs under consideration of the limited capacity of the RIS control link and practical reflection coefficients, as $N$ increases to infinity.
As proved in \cite{ref.Wu2018beamforming}, given an ideal RIS that reflects the incident wave without power loss under unlimited control link capacity, the downlink SNR of the RIS achieves, asymptotically, the order of $\mathcal{O}\left(N^2\right)$, as $N$ increases to infinity.
However, the downlink SNRs of a conventional massive MIMO or MIMO relay system, each of which equipped with $N$ antennas, equally increase with $\mathcal{O}\left(N\right)$ as proved in \cite{ref.Ngo2013energy}.
This squared SNR gain of RIS will analytically result in twice as much performance as conventional array systems in terms of achievable rate, without additional radio resources.
In order to prove the optimality of the achievable rate using practical RISs under the aforementioned limitations, we first design a passive beamformer that achieves the SNR order of $\mathcal{O}\left(N^2\right)$ asymptotically.
We then design a modulation scheme which can be used in an RIS that uses ambient RF signals to transmit data without additional radio resource and achieves the asymptotic SNR in order of $\mathcal{O}\left(N\right)$ like a conventional massive MIMO (or MIMO relay) system.
We finally propose a resource allocation algorithm to maximize the sum-rate of the considered RIS-based MISO system.

\vspace{-0.5cm}
\subsection{Passive Beamformer Design}\vspace{-0.2cm}
The maximum instantaneous SNR at DUE $k$ can be achieved by using a maximum ratio transmission (MRT) where ${\boldsymbol{w}_k}={\boldsymbol{h}_k}/\left\|{\boldsymbol{h}_k}\right\|$, which yields an SNR ${\gamma_k}={PE_{{k}}^{\rm{d}}}\left\|{\boldsymbol{h}_k}\right\|^2 / N_0$. %\cite{ref.Tse2005fundamentals}.
We then formulate an optimization problem whose goal is to maximize instantaneous SNR at RUE $k$ with respect to $\boldsymbol{\Phi}_k$ and ${\boldsymbol{w}_k}$, as follows:
\begin{equation}
\hspace{-0.5cm}\mathop {\max}\limits_{\boldsymbol{\Phi}_k, {\boldsymbol{w}_k}}\ \frac{{PE_{{k}}^{\rm{r}}}}{N_0}\left|\boldsymbol{f}_k^{\rm{H}}{\boldsymbol{\Phi} _k}\boldsymbol{G}_k {\boldsymbol{w}_k}\right|^2,\vspace{-12pt} \label{eq.P0}\\
\end{equation}\addtocounter{equation}{-1}\begin{subequations}
\begin{align}
\hspace{0.7cm}{\rm{s.t.}} \ &\left| {{\Gamma}} \right|_{\rm{min}}  \le A\left(\angle \Gamma_n^k\right) \le 1, \forall  n,\\
%\end{equation}\vspace{-0.9cm}
%\begin{equation}
&-\pi  \le \angle {\Gamma _n^k} \le \pi, \forall  n.
\end{align}
\end{subequations}
%This problem is non-convex since $\left|\boldsymbol{f}_k^{\rm{H}}{\boldsymbol{\Phi} _k}\boldsymbol{G}_k {\boldsymbol{w}_k}\right|$ is not jointly concave with respect to $\boldsymbol{\Phi}_k$ and ${\boldsymbol{w}_k}$.
For any given $\boldsymbol{\Phi}_k$, it is well known that the MRT precoder is the optimal solution to problem (\ref{eq.P0}) such that $\boldsymbol{w}_k =\frac{ \boldsymbol{G}_k^{\rm{H}}{\boldsymbol{\Phi} _k^{\rm{H}}}  \boldsymbol{f}_k}{\left\|  \boldsymbol{G}_k^{\rm{H}}{\boldsymbol{\Phi} _k^{\rm{H}}}  \boldsymbol{f}_k \right\|}$
 \cite{ref.Tse2005fundamentals}.
\begin{algorithm}
\caption{Reflection Phase Selection Algorithm}
\label{a.1}
\begin{algorithmic}[1]
\State \scriptsize{\textbf{Initialization:}} Select $\hat m = \mathop {\arg \max }\limits_{1 \le m \le M}  \left\| {{{\boldsymbol{g}}_m^k}} \right\|$ and set $s_{0}=0$ and $i=1$.
%\State \textbf{Reflecting element selection:} ${\hat n} = \mathop {\arg \max }\limits_{n \in \mathcal{N}} \left| {{s_{{{i - 1}}}} + a_{n\hat m} \phi \left( { -\pi} \right)}\right|$.%+ \left| {{f_{\tilde n}}} \right|\left| {{g_{\tilde n\hat m}}} \right|{e^{j\left( { - \pi  + \angle f_{\tilde n}^* + \angle {g_{\tilde n\hat m}}} \right)}}} \right)$.
\State \textbf{Reflection phase selection:} $\hat{\theta} = \mathop {\arg \max }\limits_{{\theta} \in \mathcal{P}} \left| {{s_{{{i - 1}}}} + a_{i,\hat m}^k \phi \left( { \theta} \right)}\right|$.
\State \textbf{Update reference value:} ${s_i} ={s_{i-1}} +   a_{i,\hat m}^k \phi \big( { \hat\theta} \big)$.
\State Select ${\phi _{{i}}} =  \phi \big( { \hat\theta} \big)$.
\State Set $i\leftarrow i + 1$ and go to Step 2 until $i=N+1$.
\State Return ${{\boldsymbol{\hat \Phi }}_k} = {\rm{diag}}\left( {{\phi _1},{\phi _2}, \cdots ,{\phi _N}} \right)$.
\end{algorithmic}
\end{algorithm}
%To solve (\ref{eq.P0}), we propose an alternating optimization algorithm that can achieve, asymptotically, $\mathcal{O}\left(N^2\right)$ as $N \to \infty$. 
%To solve (\ref{eq.P0}), 
Then, we formulate an optimization problem with respect to $\boldsymbol{\Phi}_k$ as follows: %under equal power allocation, as follows:
\begin{equation}
\hspace{-1.1cm}\mathop {\max}\limits_{\boldsymbol{\Phi}_k} \  \frac{PE^{\rm{r}}_k}{N_0}{\left\|\boldsymbol{f}_k^{\rm{H}}{\boldsymbol{\Phi} _k}\boldsymbol{G}_k \right\|^2},\vspace{-14pt} \label{eq.P1}\\
\end{equation}\addtocounter{equation}{-1}\begin{subequations}
\begin{align}
\hspace{-1.5cm}{\rm{s.t.}} \ & \text{(8a), (8b).} \nonumber
\end{align}
\end{subequations}
%\begin{align}
%\hspace{0.5cm}{\rm{s.t.}} \ & \left| {{\Gamma}} \right|_{\rm{min}}  \le A\left(\angle \Gamma_n\right)\le 1, \forall  n,\\
%%\end{equation}\vspace{-0.9cm}
%%\begin{equation}
% &-\pi  \le\angle {\Gamma _n} \le \pi, \forall  n.
%\end{align}
%\end{subequations}
This problem is non-convex since $\left\|\boldsymbol{f}_k^{\rm{H}}{\boldsymbol{\Phi} _k}\boldsymbol{G}_k \right\|^2$ is not concave with respect to $\boldsymbol{\Phi}_k$.
Let ${{\boldsymbol{f}}_k} = \left[ {{f_1^k}, \cdots ,{f_N^k}} \right]^{\rm{T}}$ and ${{\boldsymbol{G}}_k} = \left[ {{{\boldsymbol{g}}_1^k}, \cdots {{\boldsymbol{g}}_M^k}} \right]$, where ${{\boldsymbol{g}}_m^k} \in {\mathbb{C}^{N \times 1}}=\left[ {{g_{1,m}^k}, \cdots ,{g_{N,m}^k}} \right]^{\rm{T}}$ is the channel between BS antenna $m$ and RIS $k$.
Then, ${\left\| {{\boldsymbol{f}}_k^{\rm{H}}{{\boldsymbol{\Phi }}_k}{{\boldsymbol{G}}_k}} \right\|^2}$ is obtained by
\begin{equation}
{\left\| {{\boldsymbol{f}}_k^{\rm{H}}{{\boldsymbol{\Phi }}_k}{{\boldsymbol{G}}_k}} \right\|^2} = {\sum\limits_{m = 1}^M {\left| {\sum\limits_{n = 1}^N {\left| {{f_n^k}} \right|\left| {{g_{n,m}^k}} \right|A\left(\angle \Gamma_n^k\right){e^{j\left( {\angle {\Gamma _n^k} + \angle f_n^{k*} + \angle {g_{n,m}^k}} \right)}}} } \right|} ^2}
 = {\sum\limits_{m = 1}^M {\left| {\sum\limits_{n = 1}^N {a_{n,m}^k}\cdot \phi \left( {\angle {\Gamma _n^k}} \right) } \right|} ^2},
\label{eq.AO}
\end{equation}
where ${a_{n,m}^k} = \left| {{f_n^k}} \right|\left| {{g_{n,m}^k}} \right|{e^{j\left( {\angle f_n^{k*} + \angle {g_{n,m}^k}} \right)}}$ and $\phi \left( {\angle {\Gamma _n^k}} \right) = A\left(\angle \Gamma_n^k\right){e^{j\angle {\Gamma _n^k}}}$.
Since $ \left| {{f_n^k}} \right|\left| {{g_{n,m}^k}} \right|A\left(\angle \Gamma_n^k\right)$ is always greater than zero given that $\left| {{\Gamma}} \right|_{\rm{min}}>0$, (\ref{eq.AO}) can readily achieve $\mathcal{O}\left(N^2\right)$ when $\angle {\Gamma _{n}^k} = - \angle f_n^{k*} - \angle {g_{n,m_0}^k}$ for $1   \le m_0 \le M$ and $\forall n$.
However, using continuous reflection phases is impractical when we have a limited RIS control link capacity and practical RIS hardware. %implementation. %, as discussed in \cite{ref.Wu2019bfoptimization}.
%Moreover, a reflection amplitude, $\left|{\Gamma_n}\right|$, depends on the value of $\angle{\Gamma_n}$ and should be considered in the phase selection to maximize the instantaneous SNR.
Given a discrete reflection phase set $\mathcal{P}=\{- \pi, - \pi + \Delta \phi,  \cdots, - \pi + \Delta \phi(2^b -1) \}$ where $\Delta \phi = 2\pi / 2^b$ and $b$ is the number of RIS control bits at each reflecting element,
we design a suboptimal reflection matrix ${\boldsymbol{\hat \Phi} _k}$ which can achieve $\mathcal{O}\left(N^2\right)$ as $N$ increases, asymptotically, as shown in Algorithm {\ref{a.1}}.
In this algorithm, we first select antenna $\hat m$ which has the largest channel gain among $M$ channels between the BS and the RIS.
%and also select reflecting element $\hat n$ whose effective channel, $a_{\hat n \hat m}$, has the maximum gain among $N$ channels at antenna $\hat m$.
Since a reflection amplitude is always 1 when its phase equals to $-\pi$, 
%we have $\hat \theta = -\pi$ in Step 3 and 
%$\phi \left( { -\pi} \right)$ is determined as a reflection phase at $\hat n$ 
%and we select $a_{\hat n\hat m} \phi \left( { -\pi} \right)$ as a reference vector, $s_1$, in the first round.
$\phi \left( { -\pi} \right)$ is selected as a reflection phase $\phi_1$
and we determine $a_{1,\hat m} \phi \left( { -\pi} \right)$ as a reference value, $s_1$, in the first round.
%Note that the reflection coefficient decreases from $1$ to $\left|\Gamma\right|_{\rm{min}}$ as its phase increases from $-\pi$ to 0, respectively.
%When $i>1$, we compare the Euclidean norm of vector additions between $s_{i-1}$ and $-\pi$ shifted candidates, i.e., $a_{n\hat m} \phi \left( { -\pi} \right)$,
%and select $\hat n$ with the maximum Euclidean norm.
%By comparing this Euclidean norm at $\hat n$ in terms of $\theta \in \mathcal{P}$, we can select $\hat \theta$ and derive suboptimal solution such that ${\phi _{{\hat n}}} =  \phi \big( { \hat\theta} \big)$.
%Since we consider the two-stage selection of the reflecting element and reflection phase, such as Step 2 and 3 in Algorithm {\ref{a.1}}, this algorithm will not achieve the optimal performance that can be obtained by the exhaustive search method.
Note that ${\left\| {{\boldsymbol{f}}_k^{\rm{H}}{{\boldsymbol{\Phi }}_k}{{\boldsymbol{G}}_k}} \right\|^2}$ is calculated based on the sum of $N$ vectors such as ${\sum\nolimits_{n = 1}^N {a_{n,m}^k} \phi \left( {\angle {\Gamma _n^k}} \right) }$, as shown in (\ref{eq.AO}).
Therefore, when $i>1$, we compare the Euclidean norm of vector additions between $s_{i-1}$ and $\theta$ shifted candidates, i.e., $\left| {{s_{{{i - 1}}}} + a_{i,\hat m}^k \phi \left( { \theta} \right)}\right|, \forall \theta \in \mathcal{P}$,
and select $\hat \theta$ with the maximum Euclidean norm.
Therefore, we can derive the suboptimal solution such that ${\phi _{{i}}} =  \phi \big( { \hat\theta} \big)$ for each reflecting element $i$.
Algorithm {\ref{a.1}} results in a suboptimal solution and will not achieve the optimal performance that can be obtained by the exhaustive search method with $\mathcal{O}{\left({2^{bN}} \right)}$ complexity.
However, we can prove the following result related to the asymptotic optimality of Algorithm {\ref{a.1}}.

{\bf{{Proposition 1.}}} Algorithm {\ref{a.1}} can achieve an instantaneous SNR in order of $\mathcal{O}\left(N^2\right)$ regardless of the number of RIS control bits, $b\ge 1$, with a complexity of  $\mathcal{O}\left(N \right)$,
as $N$ increases.
\begin{proof}
In order to analyze the impact of $b$ on the instantaneous SNR resulting from Algorithm {\ref{a.1}}, we first consider the case of $b=1$ with $\mathcal{P}=\{- \pi, 0 \}$.
In Step 2 of Algorithm {\ref{a.1}}, the Euclidean norm of vector addition is obtained as follows:
\begin{align}
\left| {{s_{i - 1}} + {a_{i,\hat m}^k}\phi \left( \theta  \right)} \right| &= \sqrt {{{\left| {{s_{i - 1}}} \right|}^2} + {{\left| {{a_{i,\hat m}^k}\phi \left( \theta  \right)} \right|}^2} + 2\left| {{s_{i - 1}}} \right|\left| {{a_{i,\hat m}^k}\phi \left( \theta  \right)} \right|\cos \delta } \\
&\mathop  \ge \limits_{\left( a \right)} \sqrt {{{\left| {{s_{i - 1}}} \right|}^2} + {{\left| {{f_{i}^k}} \right|}^2}{{\left| {{g_{i,\hat m}^k}} \right|}^2}\left| \Gamma  \right|_{{\rm{min}}}^2 + 2\left| {{s_{i - 1}}} \right|\left| {{f_{i}^k}} \right|\left| {{g_{i,\hat m}^k}} \right|{{\left| \Gamma  \right|}_{{\rm{min}}}}\cos \delta }, \label{eq.lemma1lb}
\end{align}
where $\delta  = \angle {s_{i - 1}} - \angle {a_{i,\hat m}^k} - \theta $ and (a) results from the worst case scenario where $\left| {\phi \left( \theta  \right)}  \right| = \left| \Gamma  \right|_{\rm{min}}$.
Given that $\theta \in \{- \pi, 0 \}$ and $\left| {{\Gamma}} \right|_{\rm{min}}>0$, we can always select $ \theta$ that satisfies $\cos \delta \ge 0$ and then, $\left|s_i\right|$ will increase as $i$ increases until $i = N$.
Therefore, $\left|s_N\right|$ will increase with $\mathcal{O}\left(N\right)$ as $N$ increases.
Since ${\left| {\sum\nolimits_{n = 1}^N {a_{n,\hat m}^k}\cdot \phi \left( {\angle {\Gamma _n^k}} \right) } \right|}^2$ in (\ref{eq.AO}) equals to $\left|s_N\right|^2$ in Algorithm {\ref{a.1}},
${\left\| {{\boldsymbol{f}}_k^{\rm{H}}{{\boldsymbol{\hat\Phi }}_k}{{\boldsymbol{G}}_k}} \right\|^2}$ increases with $\mathcal{O}\left(N^2\right)$ as $N$ increases.
Also, the complexity of Algorithm 1 depends on the calculation of the maximum Euclidean norm.
Since Algorithm 1 compares $\left| {{s_{{{i - 1}}}} + a_{i,\hat m}^k \phi \left( { \theta} \right)}\right|$ for all ${{\theta} \in \mathcal{P}}$ in Step 2,
the complexity at each round $i$ will be $\mathcal{O}\left(2^{b+1} \right)$.
Hence, the total complexity of Algorithm 1 will be $\mathcal{O}\left(2^{b+1}N \right)$
and will increase with $\mathcal{O}\left(N \right)$ as $N$ increases, based on the scaling law for $N$.
Therefore, Algorithm 1 asymptotically requires a complexity of $\mathcal{O}\left(N\right)$.
\end{proof}

Proposition 1 shows that the instantaneous SNR resulting from Algorithm {\ref{a.1}} can be in order of $\mathcal{O}\left(N^2\right)$ even with one bit control for each reflecting element.
Proposition 1 also shows that Algorithm {\ref{a.1}} can achieve an instantaneous SNR in the order of $\mathcal{O}\left(N^2\right)$ for all complex channels.
Algorithm {\ref{a.1}} can be applied to the practical scenario in which UEs receive signals from multiple different RISs.
In this case, Algorithm 1 can be applied according to each RIS channel,
and it can also achieve an instantaneous SNR in the order of $\mathcal{O}\left(N^2\right)$.
Moreover, Algorithm {\ref{a.1} asymptotically requires a complexity of $\mathcal{O}\left(N\right)$ resulting in simpler RIS control compared to the various existing works on RIS in \cite{ref.Wu2019bfoptimization,ref.Wu2018beamforming,ref.Abey2019intelligent}, and \cite{ref.MMZ}.
Specifically, Algorithm 1 asymptotically requires a complexity of $\mathcal{O}\left(N\right)$ that is much lower than the complexity of baselines such as the algorithm in \cite{ref.Abey2019intelligent} whose complexity is $\mathcal{O}\left(N^2\right)$.
Since we consider an asymptotic optimality for a large $N$, this is a significant difference and advantage of our work compared to the prior art.
%Next, we derive the optimal transmit precoding vector, $\boldsymbol{w}_k$, under consideration of the RIS reflection matrix from Algorithm {\ref{a.1}}.

Next, we analyze the average SNR of a downlink RIS system, under consideration of the RIS reflection matrix derived from Algorithm {\ref{a.1}}.
We first consider an ideal RIS control link that can use an infinite and continuous reflection phases at the RIS.
%and assume that $\boldsymbol{f}_k  \sim \mathcal{CN}\left(0,\boldsymbol{I}_N\right)$ and $\boldsymbol{g}_m  \sim \mathcal{CN}\left(0,\boldsymbol{I}_M\right)$ considering Rayleigh fading channels.
From (\ref{eq.AO}), the instantaneous SNR at RUE $k$ is obtained by
\begin{equation}
\gamma_k =\frac{{{PE_{{k}}^{\rm{r}}}}}{N_0} {\left\| {{\boldsymbol{f}}_k^{\rm{H}}{{\boldsymbol{\Phi }}_k}{{\boldsymbol{G}}_k}} \right\|^2}
= \frac{{{PE_{{k}}^{\rm{r}}}}}{N_0}{\sum\limits_{m = 1}^M {\left| {\sum\limits_{n = 1}^N {\left| {{f_n^k}} \right|\left| {{g_{n,m}^k}} \right|A\left(\angle \Gamma_n^k\right){e^{j\left( {\angle {\Gamma _n^k} + \angle f_n^{k*} + \angle {g_{n,m}^k}} \right)}}} } \right|} ^2}.\label{eq.gamma_k2}
\end{equation}
By selecting ${\angle {\Gamma _n^k} = \theta_n^k =  -\angle f_n^{k*} - \angle {g_{n,m_0}^k}} $ for $0 \le m_0 \le M$ and $\forall n$, we have the following:
\begin{align}
\frac{{{PE_{{k}}^{\rm{r}}}}}{N_0}{\left\| {{\boldsymbol{f}}_k^{\rm{H}}{{\boldsymbol{\hat \Phi }}_k}{{\boldsymbol{G}}_k}} \right\|^2} &\ge \frac{{{PE_{{k}}^{\rm{r}}}}}{N_0} \left({\left| {\sum\limits_{n = 1}^N {\left| {{f_n^k}} \right|\left| {{g_{n ,m_0}^k}} \right|A\left(\theta_n^k\right) }}\right| ^2}
+\sum\limits_{m\ne m_0}^M {\left| {\sum\limits_{n = 1}^N {\left| {{f_n^k}} \right|\left| {{g_{n ,m}^k}} \right|A\left(\theta_n^k\right) {e^{j\Delta {g_{n,m,{m_0}}^k}}}}}\right| ^2}\right)\nonumber \\
&\mathop  \ge \limits_{\left( b \right)}  \frac{{{PE_{{k}}^{\rm{r}}}}\left| \Gamma \right|_{\rm{min}}^2}{N_0}\left({\left| {\sum\limits_{n = 1}^N {\left| {{f_n^k}} \right|\left| {{g_{n,m_0}^k}} \right|  }}\right| ^2 } +{\sum\limits_{m\ne m_0}^M {\left| {\sum\limits_{n = 1}^N {\left| {{f_n^k}} \right|\left| {{g_{n,m}^k}} \right|{e^{j\Delta {g_{n,m,{m_0}}^k}}}}}\right| ^2}}\right), \label{eq.SNRlb}
\end{align}
where $\Delta {g_{n,m,{m_0}}^k}=\angle g_{n,m}^k -\angle g_{n,m_0}^k$ and $\boldsymbol{\hat \Phi}_k$ and (b) result from Algorithm {\ref{a.1}} with infinite $b$ and the minimum reflection amplitude, i.e., $A\left( {{\theta _n^k}} \right) \ge {\left| \Gamma  \right|_{\min }}$, $\forall n$, respectively.
We refer to (\ref{eq.SNRlb}) as an instantaneous SNR lower bound $\munderbar{\gamma}_k$.
For notational convenience, we define $\munderbar \gamma_{\rm{l}}^k = {\sum\nolimits_{n = 1}^N {\left| {{f_n^k}} \right|\left| {{g_{n,m_0}^k}} \right|  }}$ and $\munderbar \gamma_{{\rm{r}}m}^k ={ {\sum\nolimits_{n = 1}^N {\left| {{f_n^k}} \right|\left| {{g_{n,m}^k}} \right|{e^{j\Delta {g_{n,m,{m_0}}^k}}}}}}$ in (\ref{eq.SNRlb}).
Then, the random variable $\munderbar \gamma_k$ follows Lemma 1.

{\bf{{Lemma 1.}}} Based on the scaling law for $N$, the mean of $\munderbar \gamma_k$ has a lower bound which increases with $\mathcal{O}\left(N^2\right)$ as $N$ increases, as follows:
\begin{equation}
E\big[ {{{\munderbar \gamma }_k}} \big] \ge \frac{{{PE_{{k}}^{\rm{r}}}}\left| \Gamma \right|_{\rm{min}}^2}{N_0}{\left( {\sum\limits_{n= 1}^N {\bar\mu _{k,n}^{\rm{b}}\bar\mu _{k,n}^{\rm{r}}} } \right)^2},
\end{equation}
where
\begin{align}
\bar\mu _{k,n}^{\rm{b}} &= \sqrt {\frac{\pi }{4  \left({{{{\kappa_{{\rm{b}}k}} + 1}}}\right)}\sum\nolimits_{j = 1}^N {\left| {{{\left( {{{\boldsymbol{R}}_{{\rm{b}}k}}} \right)}_{n,j}}} \right|} } {L_{{{\frac{1}{ 2}}}}}\left( { - \frac{{{{{\kappa_{{\rm{b}}k}}\left| {\bar g_{n,m_0}^k} \right|}^2}}}{{\sum\nolimits_{j = 1}^N {\left| {{{\left( {{{\boldsymbol{R}}_{{\rm{b}}k}}} \right)}_{n,j}}} \right|} }}} \right),\\
\bar\mu _{k,n}^{\rm{r}} &= \sqrt {\frac{\pi }{4\left({\kappa_{{\rm{r}}k}}+1\right)}\sum\nolimits_{j = 1}^N {\left| {{{\left( {{{\boldsymbol{R}}_{{\rm{r}}k}}} \right)}_{n,j}}} \right|} } {L_{{{\frac{1}{ 2}}}}}\left( { - \frac{{{{{\kappa_{{\rm{r}}k}}\left| {\bar f_n^k} \right|}^2}}}{{\sum\nolimits_{j = 1}^N {\left| {{{\left( {{{\boldsymbol{R}}_{{\rm{r}}k}}} \right)}_{n,j}}} \right|} }}} \right),
\end{align}
where $\boldsymbol{}$$L_n \left(x\right)$ is the Laguerre polynomial of order $n$, $\bar g^k_{n,m_0} = \left({\boldsymbol{\bar G}_k}\right)_{n,m_0}$, and ${{\boldsymbol{\bar f}}_k} = \left[ {{\bar f^k_1}, \cdots ,{\bar f^k_N}} \right]^{\rm{T}}$.
\begin{proof}
The detailed proof is presented in Appendix A.
\end{proof}}
Lemma 1 shows that the lower bound of the average SNR increases with $\mathcal{O}\left(N^2\right)$ and the average SNR gains from $M-1$ antennas become negligible compared to those from $m_0$, as $N$ increases.
Moreover, we can observe that this lower bound is equal to the single antenna case in \cite{ref.Basar2019wireless}.
Since $\mathcal{O}\left(N^2\right)$ can also be achieved by the instantaneous SNR resulting from Algorithm {\ref{a.1}} with limited $b$, as proved in Proposition 1,
the average SNR gain from the antenna $m_0$ can asymptotically achieve the instantaneous SNR resulting from Algorithm {\ref{a.1}}
and the average SNR gains from $M-1$ antennas are also negligible compared to those from $m_0$ in the limited RIS control link capacity, as $N$ increases.
Therefore, the average SNR resulting from Algorithm {\ref{a.1}} with limited $b$ will also converge to that of the $m_0$-th transmit antenna selection.
%Therefore, the average SNR resulting from Algorithm {\ref{a.1}} with limited $b$ also decreases as $M$ increases and the maximum SNR can be achieved when $M=1$.
%Thus, the antenna selection can eventually achieve this maximum SNR.
Given this convergence of the average SNR, we can use $\boldsymbol{w}_k$ as an antenna selection that can achieve full multi-antenna diversity with a low-cost and low-complexity instead of a MRT \cite{ref.Sanayei2004antenna}.
By selecting the BS antenna whose channel gain has the maximum value such as in Step 1 of Algorithm {\ref{a.1}}, we can determine the transmit precoding vector, ${{\boldsymbol{w}}_k} = {\left[ {{w_1^k}, \cdots ,{w_M^k}} \right]^{\rm{T}}}$, as follows:
\begin{equation}
\left\{ \begin{array}{l}
{w_m^k} = 1,\ {\text{if }} m = \hat m_k,\\
{w_m^k} = 0,\ {\text{if }} m \ne \hat m_k,
\end{array} \right.\label{eq.AS}
\end{equation}
where $\hat m_k = \mathop {\arg \max }\limits_{1 \le m \le M} \left\| {{{\boldsymbol{g}}_m^k}} \right\|$.
Although an MRT precoder achieves the optimal performance for a single-user MISO system, it requires multiple RF chains associated with multiple antennas resulting in higher cost and hardware complexity compared to the transmit antenna selection scheme.
Moreover, since the average SNR will converge to the single antenna system as $N$ increases, a large $N$ results in a performance convergence between MRT and transmit antenna selection.
In fact, our analysis is viewed as a generalized version of \cite{ref.Wu2019bfoptimization} and \cite{ref.Abey2019intelligent}.
We have jointly considered practical RIS environments and their limitations, such as practical reflection coefficients (i.e., reflection power loss) and discrete phase shifts,
and we proved that  Algorithm 1 can achieve a SNR in order of $\mathcal{O}\left(N^2\right)$ regardless of the number of RIS control bits.
However, in \cite{ref.Wu2019bfoptimization}, the authors considered discrete phase shifts with lossless reflection coefficients
and the authors in \cite{ref.Abey2019intelligent} considered infinite and continuous phase shifts.
Moreover, in the revised manuscript, we have proved that Algorithm 1 achieves an instantaneous SNR in order of $\mathcal{O}\left(N^2\right)$ for all complex channels regardless of the distribution of the channels and also achieves an average SNR in order of $\mathcal{O}\left(N^2\right)$ for spatially correlated Rician fading channels.
However, in \cite{ref.Wu2019bfoptimization}, the authors proved that an RIS with discrete phase shifts achieves an average SNR in order of $\mathcal{O}\left(N^2\right)$ for indenpendent  Rayleigh fading channels which is impractical assumption for an RIS with densely located reflecting elements.
Also, Algorithm 1 requires a complexity of $\mathcal{O}\left(N\right)$ that is much lower than that in \cite{ref.Abey2019intelligent} whose complexity is $\mathcal{O}\left(N^2\right)$, resulting in a significant difference and advantage of our work compared to \cite{ref.Abey2019intelligent} for a large $N$.

As a special case of Lemma 1, we can verify the average SNR for i.i.d. Rayleigh fading channels by assuming that ${\kappa_{{\rm{b}}k}}={\kappa_{{\rm{r}}k}}=0$ and ${\boldsymbol{R}_{{\rm{b}}k}}={\boldsymbol{R}_{{\rm{r}}k}}={\boldsymbol{I}}_N$.
In this case, $\left| f_n^k \right|$ and $\left| g_{n, m_0}^k\right|$ are independent random variables for different $n$.
Hence, $\munderbar \gamma_{\rm{l}}^k$ converges to a Gaussian distribution based on the central limit theorem (CLT) \cite{ref.Basar2019wireless}:
${\munderbar \gamma _{\rm{l}}^k} \sim \mathcal{N}\left( {\frac{{N\pi }}{4},N\left( {1 - \frac{{{\pi ^2}}}{{16}}} \right)} \right)$, as $N$ increases. % as proved in \cite{ref.Basar2019wireless}.
Then, $\big|{\munderbar \gamma _{\rm{l}}^k}\big|^2$ follows a non-central chi-squared distribution with one degree of freedom with mean of $N\left( {1 + \frac{{{\pi ^2}}}{{16}}\left( {N - 1} \right)} \right)$ and variance of ${N^2}\left( {1 - \frac{{{\pi ^2}}}{{16}}} \right)\left( {2 - \frac{{{\pi ^2}}}{8} + \frac{{N{\pi ^2}}}{4}} \right)$.
Similarly, $\munderbar \gamma_{{\rm{r}}m}^k$ converges to a complex Gaussian distribution as
${\munderbar \gamma _{{\rm{r}}m}^k} \sim \mathcal{CN}\left( {0,N} \right)$, and $\big|\munderbar \gamma_{{\rm{r}}m}^k\big|^2$ follows a central chi-squared distribution with two degrees of freedom with mean $N$ and variance $N^2$, as $N$ increases.
Since $\left|\bar \gamma_{{\rm{r}}m}^k\right|^2$ are independent random variables for different $m$ and also independent with $\big|{\munderbar \gamma _{\rm{l}}}^k\big|^2$,
we have the following mean and variance of $\munderbar \gamma_k$, respectively:
\begin{align}
{\rm{E}}\big[ {{{\munderbar \gamma }_k}} \big] &= \frac{{NPE_k^{\rm{r}}\left| \Gamma  \right|_{{\rm{min}}}^2}}{{{N_0}}}\left( {M + \frac{{{\pi ^2}\left( {N - 1} \right)}}{{16}}} \right),\label{eq.Egamma}\\
{\rm{Var}}\big[ {{{\munderbar \gamma }_k}} \big] &= \frac{{{N^2}{P^2}{{\left( {E_k^{\rm{r}}} \right)}^2}\left| \Gamma  \right|_{{\rm{min}}}^4}}{{N_0^2}}\left\{ {\left( {1 - \frac{{{\pi ^2}}}{{16}}} \right)\left( {2 - \frac{{{\pi ^2}}}{8} + \frac{{N{\pi ^2}}}{4}} \right) + M - 1} \right\}.\label{eq.Vgamma}
\end{align}
(\ref{eq.Vgamma}) shows that the variance of the instantanenous SNR increases with $\mathcal{O}\left(N^3\right)$, asymptotically, and this will result in scheduling diversity.
To achieve this scheduling diversity for a large $N$, we will develop a new resource allocation algorithm that can achieve the maximum scheduling diversity in Section IV. C.
Moreover, we can observe that (\ref{eq.Egamma}) with $M=1$ is equal to the average SNR derived in \cite{ref.Basar2019wireless},
showing that the SER resulting from Algorithm {\ref{a.1}} also decays exponentially as a function of $N$.

\vspace{-0.5cm}
\subsection{Modulation for Unscheduled RUE }
Next, we devolop a modulation scheme that can be used to increase the achievable RIS sum-rate.
In our downlink OFDMA scheme, the BS can transmit the downlink signal to only one scheduled UE at each time slot.
However, by reflecting the ambient RF signals generated from the BS,
each RIS also can send the downlink signal to its unscheduled RUE at each time slot,
as done in ambient backscatter communications \cite{ref.Liu2013AmbientBW}. %without additional radio resources, 
Different from ambient backscatter communications, an RIS can convey information by reflecting the ambient RF signal without interfering with existing scheduled UEs. %by using RIS characteristics.
In addition to the data rates obtained from the BS's downlink DUEs, we can obtain additional data rates at unscheduled RUEs without additional radio resources, resulting in higher achievable sum-rate than a conventional network without RIS.
For convenience, we refer to each unscheduled RUE and its connected RIS as uRUE and uRIS, respectively.
As shown in \cite{ref.Basar2019wireless} and \cite{ref.Tang2019program}, an RIS can transmit PSK signals to serving user by controlling its reflection phase.
As such, we consider that each uRIS controls its reflection phase to send the downlink signal to each corresponding uRUE by reflecting the incident wave from the BS, whenever the DUE is scheduled.
In order to avoid undesired interference from uRISs to the scheduled DUE, we consider the following procedure.
First, the BS sends the RIS control signals, which are related to the data symbols for each uRUE, to uRISs during the CSI-RS period (see Fig. {\ref{fig.8}}).
Each uRIS controls $N$ bias DC voltages for all reflecing elements depending on the control signal
and reflects the incident wave from the BS with the same reflection phase during the DMRS and data transmission periods.
Meanwhile, each scheduled DUE (sDUE) receives the DMRS from the BS and estimates the effective CSI which includes the transmit precoding at the BS and the phase shifts from the uRISs.
Since all uRISs keep using the same reflection phase from the beginning of the DMRS to the end of the data transmission, this effective CSI will not change during the downlink data transmission
and this results in zero interference.
Similarly, each uRUE receives the CSI-RS from the BS and estimates the CSI between the BS and each uRUE via corresponding uRIS without controlled phase shifts,
i.e., ${\boldsymbol{\Phi} _k}=-\boldsymbol{I}_N, \forall k \in \mathcal{R}$.
%Note that AP can share the information about the transmit precoder and the modulation of sDUE given that
%those information is included in the downlink control indicator (DCI) and the DCI is broadcasted by the AP, based on the LTE specification in \cite{ref.LTE2018TS36213} and \cite{ref.LTE2017TS36212}.
Note that the BS broadcasts the information about the transmit precoder and the modulation of sDUE by using the downlink control indicator (DCI), based on the LTE specification in \cite{ref.LTE2018TS36213} and \cite{ref.LTE2017TS36212}.
Given the estimated CSI and the broadcast DCI, the uRUE can calculate the effective CSI and eventually decode the downlink signal transmitted from the uRIS.
By using this procedure, each uRIS needs to transmit only one symbol during the channel coherence time, however, the uRUEs can achieve the asymptotic SNR in order of $\mathcal{O}\left(N\right)$ as will be proved in Theorem 1.
Moreover, since the uRUE receives the same symbols from the uRIS during the data transmission period, the uRUE will achieve a diversity gain proportional to the length of the data transmission period.
Furthermore, since the BS can send the RIS control signal related to one symbol for each uRIS during the entire channel coherence time, we can reduce the link burden of the RIS control link.
\begin{figure}[t]
\centering
\includegraphics[width=0.5\columnwidth] {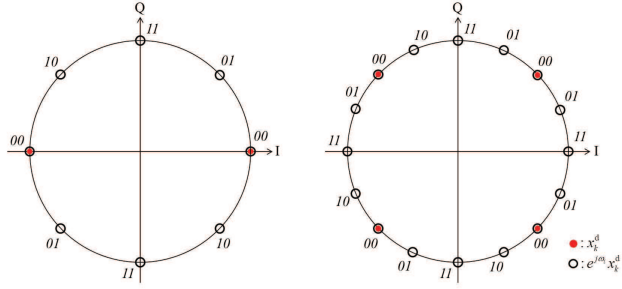}\vspace{-0.5cm}
\caption{Proposed constellation for uRUE under consideration with two examples of QPSK signaling when (left) $x_k^{\rm{d}}$ is BPSK symbol and (right) QPSK symbol.}\vspace{-0.8cm}
\label{fig.9}
\end{figure}
Given this procedure, the received downlink signal at uRUE $i$ can be obtained by:
%\begin{equation}
${y_i} = \sqrt P {\boldsymbol{f}}_i^{\rm{H}}{{\boldsymbol{\bar \Phi }}_i}{{\boldsymbol{G}}_i}{{\boldsymbol{w}}_k}x_k^{\rm{d}} + n_i^{\rm{r}}$,
%\vspace{-0.1cm}
%\end{equation}
where $i \in \mathcal{R}$, $k \in \mathcal{D}$, and ${{\boldsymbol{w}}_k}$ and $x_k^{\rm{d}}$ are the transmit precoder and the downlink signal for sDUE $k$, respectively.
To modulate the downlink data bits into the reflection matrix, we propose the following modulated reflection matrix for uRUE $i$:
\begin{equation}
{{\boldsymbol{\bar \Phi }}_i} = A\left( {{\omega _i}} \right){e^{j{\omega _i}}}{{\boldsymbol{I}}_N},\ {\omega _i} \in \mathcal{M},\label{eq.mod}\vspace{-0.1cm}
\end{equation}
where ${e^{j{\omega _i}}}$ is the proposed $M_{\rm{o}}$-PSK modulation symbol for uRUE $i$ and $\mathcal{M} = \left[ {{\mu _1}, \cdots ,{\mu _{{2^{{M_{\rm{o}}}}}}}} \right]$ is a set of the corresponding $M_{\rm{o}}$-PSK modulation.
For two examples of QPSK signaling as shown in Fig. {\ref{fig.9}}, $\mathcal{M}=\left\{0, \frac{\pi}{4}, \frac{\pi}{2}, \frac{3\pi}{4}\right\}$ when $x_k^{\rm{d}}$ is BPSK symbol
and $\mathcal{M}=\left\{0, \frac{\pi}{8}, \frac{\pi}{4}, \frac{3\pi}{8}\right\}$ when $x_k^{\rm{d}}$ is QPSK symbol.
When $x_k^{\rm{d}}$ is a QAM symbol, we can also obtain $\mathcal{M}$ according to the minimum angle between adjacent QAM symbols.
Using (\ref{eq.mod}), $y_i$ is obtained by:
\begin{equation}
{y_i} = \sqrt P {\boldsymbol{f}}_i^{\rm{H}}{{\boldsymbol{G}}_i}{{\boldsymbol{w}}_k}A\left( {{\omega _i}} \right){e^{j{\omega _i}}}x_k^{\rm{d}} + n_i^{\rm{r}}. \label{eq.yi2}
\end{equation}
Note that the channel ${\boldsymbol{f}}_i^{\rm{H}}{{\boldsymbol{G}}_i}$ can be estimated from the CSI-RS, and 
the precoding vector ${{\boldsymbol{w}}_k}$ and the modulation scheme of $x_k^{\rm{d}}$ are known at the uRUE from the broadcasted DCI.
%uRUE has knowledge about ${{\boldsymbol{w}}_k}$ and the modulation scheme of $x_k^{\rm{d}}$ from the DCI.
Assuming that all RISs are equipped with the same passive elements resulting in identical reflection coefficient models as in (\ref{eq.approx}), 
the uRUE can calculate $A\left(\omega_i\right)$ for $\left| \mathcal{M} \right|$ symbols
and eventually estimate $\omega_i$ resulting in additional data rate in the considered RIS system.
For instance, during the CSI-RS period, the BS will broadcast the common pilot signal and uRUE $i$ will estimate the actual CSI, $\boldsymbol{f}_i^{\rm{H}}\boldsymbol{G}_i$.
Next, during the DMRS period, the BS will broadcast the common pilot signal $\boldsymbol{w}_k s_{\rm{p}}$ and uRUE $i$ will estimate the effective CSI, $\boldsymbol{f}_i^{\rm{H}}{ \boldsymbol{\bar \Phi}_i}\boldsymbol{G}_i \boldsymbol{w}_k$.
Using (\ref{eq.mod}), the received signal at uRUE $i$ during the data transmission period will be:
%\begin{equation}
$y_i = \sqrt{P} A(\omega_{i})e^{j\omega_{i}}x_k^{\rm{d}}\sum\limits_{n=1}^N f_{n}^{i*}\sum\limits_{m=1}^M g_{n,m}^iw_{m}^k + n_i^{\rm{r}}.$
%\end{equation}
Since the channel $\boldsymbol{f}_i^{\rm{H}}\boldsymbol{G}_i = \left[ \sum\nolimits_{n=1}^N f_{n}^{i*} g_{n,1}^i, \cdots, \sum\nolimits_{n=1}^N f_{n}^{i*} g_{n,M}^i\right]$, the precoding vector ${{\boldsymbol{w}}_k}$, and the modulation scheme of $x_k^{\rm{d}}$ are known at the uRUE $i$,
$\sum\nolimits_{n} f_{n}^{i*}\sum\nolimits_{m} g_{n,m}^i w_{m}^k$ can be calculated at uRUE $i$ and therefore, uRUE $i$ can estimate $\omega_{i}$.
The detailed procedure of the proposed modulation scheme is provided in Fig. \ref{fig.flow}.
From (\ref{eq.yi2}), we can prove the following result related to the average SER at the uRUE.

\begin{figure}
\centering
\includegraphics[width=0.75\columnwidth] {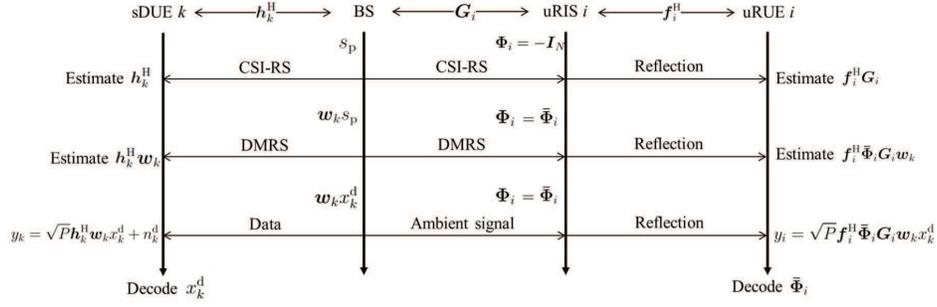}\vspace{-0.5cm}
\caption{Procedure of the proposed modulation scheme.}\vspace{-0.8cm}
\label{fig.flow}
\end{figure}

{\bf{{Theorem 1.}}} The uRUE achieves an average SNR in order of $\mathcal{O}\left(N\right)$ as $N$ increases. Also, for independent Rayleigh fading channels, the average SER with the proposed $M_{\rm{o}}$-PSK signaling can be approximated by
\begin{equation}
P_{\rm{e}}=\frac{1}{{{2^{{M_o}}}}}\sum\nolimits_{p = 1}^{{2^{{M_o}}}} {\int_0^{\pi - \frac{\Delta_\mu}{2}} {\frac{{{1}}}{{{1} - \frac{N{N_{\rm{s}}}{E_{\rm{s}}}t\left( \theta  \right)A{{\left( {{\mu _p}} \right)}^2}}{N_0}}}} d\theta },\label{eq.SER}
%\frac{1}{{{2^{{M_o}}}}}\sum\limits_{p = 1}^{{2^{{M_o}}}} {\int_0^{\frac{{\pi \left( {2{M_{\rm{o}}} - 1} \right)}}{{2{M_{\rm{o}}}}}} {\frac{{{N_0}}}{{{N_0} - N{N_{\rm{s}}}{E_{\rm{s}}}t\left( \theta  \right)A{{\left( {{\mu _p}} \right)}^2}}}} d\theta },\label{eq.SER}
\end{equation}
where
%\begin{equation}
$t\left( \theta  \right) = \frac{{ - {{\sin }^2}\left( {{{\Delta_\mu} }/{{2}}} \right)}}{{{{\sin }^2}\theta }},$
%\end{equation}
$\Delta_\mu$ is the angular spacing of the proposed $M_{\rm{o}}$-PSK symbols, and $N_{\rm{s}}$ is the number of transmitted symbols during the downlink data transmission period.
\begin{proof}
The detailed proof is presented in Appendix B.
\end{proof}
Theorem 1 shows that the uRUE can achieve an asymptotic SNR in order of $\mathcal{O}\left(N\right)$ by using the proposed modulation scheme
resulting in several implications.
First, an RIS can provide the same asymptotic SNR as a conventional massive MIMO or MIMO relay system for all unscheduled RUEs, simultaneously.
For the scheduled RUEs, an RIS also provides an asymptotic SNR in order of $\mathcal{O}\left(N^2\right)$, which is much higher than that of conventional MIMO systems, as proved in Proposition 1.
Hence, an RIS can support the demand for massive connectivity and high data traffic, without additional radio resources.
Moreover, Theorem 1 shows that the average SER of uRUE can be obtained based on deterministic values
and we can evaluate the reliability of the considered RIS system without extensive simulations.
In particular, we can observe from (\ref{eq.SER}) that the average SER decreases as $N$ increases and it eventually reaches zero as $N \to \infty$, %even in the low $E_{\rm{s}}/N_0$ region,
resulting in reliable communication regardless of $E_{\rm{s}}/N_0$ even at the uRUEs.

%Hence, the achievable sum-rate will increase proportional to $N$ and the number of RUEs without additional radio resources, supporting the demand for massive connectivity and high data traffic.
%
%Theorem 1 shows that the considered RIS system can obtain data rate at uRUEs in addition to the scheduled UE at the same time slot.
%Theorem 1 shows that the average SNR at the uRUE increases with $\mathcal{O}\left(N \right)$, asymptotically, resulting in the same SNR order of a conventional massive MIMO and MIMO relay system.
%Also, the SER decreases as $N$ increases and it eventually reaches zero as $N \to \infty$ even in the low $E_{\rm{s}}/N_0$ region.
%Moreover, the uRUE achieves a diversity order of $N_{\rm{s}}$ given that the uRUE receives the same $N_{\rm{s}}$ symbols during the data transmission period.

\vspace{-0.5cm}
\subsection{Resource Allocation Algorithm }\vspace{-0.2cm}
Next, we develop a new resource allocation algorithm for RIS based on the approaches of Section IV.
%Since the variance of the SNR increases with $\mathcal{O}\left(N^3\right)$ by the proposed phase selection algorithm, it is imperative to study the scheduling algorithm achieving those user selection diversity.
We propose a resource allocation algorithm that includes joint transmission power control and user scheduling for the maximum average sum-rate. %, as $N$ increases without bound.
Given our downlink OFDMA system, we allow the channel states to vary over $f$ and time slot $t$, in which each RB has duration of the channel coherence bandwidth and time.
In accordance with the LTE specification, the minimum scheduling period is equal to one transmission time interval (TTI) and lasts for 1 ms duration \cite{ref.LTE2017TR36211}.
For analytical simplicity, we assume that the channel coherence time is equal to one TTI
and consider a scheduling indicator $q_k^{t,f}$ that $q_k^{t,f} = 1$ when UE $k$ is scheduled in time slot $t$ and RB $f$, otherwise $q_k^{t,f} = 0$, resulting in $\sum\nolimits_{k \in \mathcal{K}}q_k^{t,f} = 1$.
Given the proposed phase selection algorithm and the modulation scheme for uRUE, 
the instantaneous achievable rate at UE $k$ in time slot $t$ is given as follows:
\begin{equation}
R_k^{t} = \left\{ {\begin{array}{*{20}{l}}
{\sum\limits_{f \in \mathcal{F}}q_k^{t,f}\log \left( {1 + \frac{{{P^{t,f}}{{\left\| {{\boldsymbol{h}}_k^{t,f}} \right\|}^2}}}{{{N_0}}}} \right), {\text{if}}\;k \in {\cal D},}\\
{\sum\limits_{f \in \mathcal{F}}\left\{q_k^{t,f}\log \left( {1 + \frac{{{P^{t,f}}{{\left| {{{\left( {{\boldsymbol{f}}_k^{t,f}} \right)}^{\rm{H}}}{\boldsymbol{\hat \Phi }}_k^{{t,f}}{\boldsymbol{\hat g}}_k^{{t,f}}} \right|}^2}}}{{{N_0}}}} \right) + \sum\limits_{i \in {\cal D}} {q_i^{t,f}} \log \left( {1 + \bar \gamma _{{\rm{a}}k}^{t,f}} \right)\right\}, {\text{if}}\;k \in {\cal R},}
\end{array}} \right. \label{eq.Rtk0}
\end{equation}
where we use superscripts $t$ and $f$ for all time-frequency-varying variables and we assume that $E_k^{\rm{r}}=E_k^{\rm{d}}=1$, $\forall k \in \mathcal{K}$.
Also, $\boldsymbol{\hat g}_k^{t,f}$ is the selected channel from $\boldsymbol{G}_k^{t,f}$ according to (\ref{eq.AS})
and ${{\bar \gamma }_{{\rm{a}}k}^{t,f}}$ is the received signal-to-interference-plus-noise ratio (SINR) at uRUE $k$ in time slot $t$ and RB $f$, which corresponds to the additional rate achieved at the uRUE, as given by:
\begin{equation}
{{\bar \gamma }_{{\rm{a}}k}^{t,f}} =   {\frac{{P^{t,f}{\alpha _i^{t,f}}\left| {\left({\boldsymbol{f}}_k^{t,f}\right)^{\rm{H}}{{\boldsymbol{\bar\Phi }}_k^{t,f}}{{\boldsymbol{G}}_k^{t,f}}{{\boldsymbol{h}}_i^{t,f}}} \right|^2}}{{\frac{{{N_0}}}{{{N_{\rm{s}}}}}{\left\|\boldsymbol{h}_i^{t,f}\right\|^2} + \sum\limits_{j \ne k,j \in \mathcal{R}} {P^{t,f}\left| {\left({\boldsymbol{f}}_{jk}\right)^{\rm{H}}{{\boldsymbol{\bar\Phi }}_j^{t,f}}{{\boldsymbol{G}}_j^{t,f}}{{\boldsymbol{h}}_i^{t,f}}} \right|^2} }}}, \label{eq.add_SINR} 
\end{equation}
where ${{\boldsymbol{f}}_{jk}^{t,f}}$ is the interference channel between uRIS $j$ and uRUE $k$,
and ${{\alpha _i^{t,f}}}$ denotes the SNR loss at uRUE resulting from the modulation order at sDUE $i$.
As the modulation order at the sDUE increases, $\Delta_\mu$ in (\ref{eq.SER}) decreases resulting in a throughput loss at the uRUE and $\alpha_i^{t,f}$ captures this throughput loss.
Since the CQI of DUE $i$ is higher than the pre-determined threshold, we assume that $\alpha_i^{t,f} \ll  1$, $\forall i \in \cal{D}$, resulting from a high order modulation at sDUE $i$.
Given that we consider an OFDMA system and uRISs will not interfere with scheduled DUEs,
we do not consider the interference from uRISs to all scheduled UEs as seen from (\ref{eq.Rtk0}).
However, since uRUEs will experience interference resulting from the undesired reflection wave generated by neighboring uRISs,
we consider the interference from other uRISs to uRUE as in (\ref{eq.add_SINR}).
Then, the instantaneous sum-rate in time slot $t$ can be obtained as follows:
\begin{equation}
{R^{t}} = \sum\limits_{f \in \mathcal{F}}\left[\sum\limits_{i \in \mathcal{D}} q_i^{t,f} \left\{\log \left( {1 +\bar \gamma_{{\rm{d}}i}^{t,f} } \right) + \sum\limits_{k \in \mathcal{R}} {\log \left( {1 + {{\bar \gamma }_{{\rm{a}}k}^{t,f}}} \right)}\right\} + \sum\limits_{k \in \mathcal{R}} q_k^{t,f} \log \left( {1 + \bar \gamma_{{\rm{r}}k}^{t,f}} \right)\right],\label{eq.SR}
\end{equation}
where $ \bar \gamma_{{\rm{d}}i}^{t,f}=\frac{{P^{t,f}}\left\|{\boldsymbol{h}_i^{t,f}}\right\|^2}{N_0}$ and $\bar \gamma_{{\rm{r}}k}^{t,f} =\frac{{P^{t,f}}\left|\left(\boldsymbol{f}_k^{t,f}\right)^{\rm{H}}{\boldsymbol{\hat \Phi} _k^{t,f}}\boldsymbol{\hat g}_k^{t,f} \right|^2}{N_0}$.
Hence, the average sum-rate of the considered RIS-based MISO system will be:
%\begin{equation}
$R = \sum\limits_{s_t \in \mathcal{S}} {{\pi _t}}  {{R^{t}}},$
%\end{equation}
where the set $\mathcal{S}$ consists of $S$ system states and each system state represents one of the possible channel states for all links \cite{ref.Lee2006oppor}.
Also, $s_t $ is the system state at time slot $t$ and $\pi_t$ is the probability that the actual system state is in state $s_t$, $\forall s_t \in \mathcal{S}$.
In accordance with the LTE specification \cite{ref.LTE2017TS36521}, a CSI is the quantized information which consists of the limited feedback bits 
and discrete system states, $\mathcal{S}$, is known at the BS which include all possible quantized CSI (i.e., $s_t \in \mathcal{S}$).
%Then, the BS can estimate the actual system state from the CSI feedback by comparing 
% the CSI feedback by comparing given 
%(i.e., $s_t \in \mathcal{S}$).
%receives the actual system state from the feedback signals
%
% the actual system state from the CSI feedback 
%Therefore, the BS has discrete system states, $\mathcal{S}$, and receives the actual system state from the feedback signals which will be one of those discrete system states (i.e., $s_t \in \mathcal{S}$).
Moreover, the number of those system states will increase as the number of feedback bits and the number of wireless links increase.
Since we consider a large $N$, we will analyze the average sum-rate of the considered RIS system as $S \to \infty$.
%For most practical communication systems, we consider a finite $S$ (i.e., finite number of time slots).
%As $S \to \infty$, the set $\mathcal{S}$ includes all possible system states, so that (\ref{eq.ER}) becomes the average sum-rate.
%In order to analyze the system performance theoretically, we first derive the resource allocation algorithm that maximizes the average sum-rate (i.e., $S \to \infty$)
%and show that the proposed algorithm also achieves the optimality in terms of the average sum-rate under the limited communication time (i.e., finite $S$).

Next, we formulate an optimization problem whose goal is to maximize the average sum-rate with respect to the scheduling indicator (i.e., $\boldsymbol{q} = \left[ \boldsymbol{\bar q}^1, \cdots, \boldsymbol{\bar q}^S\right]$ where $\boldsymbol{\bar q}^{s_t} =\boldsymbol{q}^{t}= [ q_k^{t,f}]_{k \in \mathcal{K}, f \in \mathcal{F}}$) and the transmission power at the BS (i.e., $\boldsymbol{p} = \left[ \boldsymbol{\bar p}^1, \cdots, \boldsymbol{\bar p}^S\right]$ where $\boldsymbol{\bar p}^{s_t} = \boldsymbol{p}^{t}= [ P^{t,f}]_{ f \in \mathcal{F}}$)
as follows:
\begin{equation}
\hspace{-4.0cm}\mathop {\max}\limits_{\boldsymbol{q}, {\boldsymbol{p}}}\ \sum\nolimits_{s_t  \in \mathcal{S}} {{\pi _t}}{{R^{t}}},\vspace{-12pt} \label{eq.P2}\\
\end{equation}\addtocounter{equation}{-1}\begin{subequations}
\begin{align}
\hspace{0.7cm}{\rm{s.t.}} \ & \sum\nolimits_{s_t  \in \mathcal{S}} {{\pi _t}}  {{R_k^{t}}} \ge \bar{R}, \forall  k \in \mathcal{K},\label{eq.ST2}\\
%\sum\nolimits_{t \in \mathcal{S}} {{\pi _t}}  {{R_k^{t,f}}} \ge \bar{R}_{\rm{d}}, \forall  k \in \mathcal{D},\label{eq.ST1}\\
%& \sum\nolimits_{t \in \mathcal{S}} {{\pi _t}}  {{R_k^{t,f}}} \ge \bar{R}, \forall  k \in \mathcal{K},\label{eq.ST2}\\
& \sum\nolimits_{s_t  \in \mathcal{S}} {{\pi _t}}  {{P^{t}}} \le \bar{P},\label{eq.ST3}\\
& 0 \le P^{t,f} \le P_{\rm{max}},\forall  f \in \mathcal{F}, \forall  s_t  \in \mathcal{S},\label{eq.ST4}\\
&  {q}_k^{t,f} \in \left\{ 0,1\right\},\forall  k \in \mathcal{K}, \forall  f \in \mathcal{F},\forall  s_t \in \mathcal{S},\label{eq.ST6}\\
& \sum\nolimits_{k \in \mathcal{K}} {q}_k^{t,f} =1,\forall  f \in \mathcal{F},\forall  s_t  \in \mathcal{S},\label{eq.ST5}
\end{align}
\end{subequations}
%where $\bar{R}_{\rm{d}}$ and $\bar{R}_{\rm{r}}$ are the minimum average rate requirements for the DUE and RUE, respectively,
where $P^t = \sum\nolimits_{f \in \mathcal{F}} P^{t,f}$,
$\bar{R}$ is the minimum average rate requirement,
and $\bar P$ and $P_{\rm{max}}$ are the maximum average and instantaneous transmission power per RB at the BS, respectively.
Note that (\ref{eq.P2}) is a mixed integer optimization problem which is generally hard to solve.
Hence, we relax $q_k^{t,f}$ into continuous value such that $0 \le q_k^{t,f} \le 1$,  $ \forall f \in \mathcal{F}$, $\forall k \in \mathcal{K}$, $\forall s_t  \in \mathcal{S}$.
Given that the objective function and constraints in (\ref{eq.P2}) are continuous after relaxation, 
we can easily show that the duality gap of (\ref{eq.P2}) becomes zero as $N \to \infty$ by using \cite[Proposition 1]{ref.Lee2006oppor}.
To formulate a dual problem, we define a Lagrangian function corresponding to the optimization problem in (\ref{eq.P2}) to solve its dual problem as follows:
\begin{equation}
L\left( {{{\boldsymbol{q}}},{{\boldsymbol{p}}},{\boldsymbol{\lambda }},\mu } \right) %&= \sum\limits_{s_t  \in {\cal S}} {{\pi _t}{R^{t}}}  + \sum\limits_{k \in {\cal K}} {{\lambda _k}\left( {\sum\limits_{s_t  \in {\cal S}} {{\pi _t}R_k^{t}}  - \bar R} \right)}  + \mu \left(  \bar P-{\sum\limits_{s_t  \in {\cal S}} {{\pi _t}{P^{t}}}} \right)\\
= \sum\limits_{s_t  \in \cal S}{\pi _t} L^{t} \left(\boldsymbol{q}^{t}, \boldsymbol{p}^{t},\boldsymbol{\lambda},\mu \right)  - \sum\limits_{k \in \cal K} {{\lambda _k}\bar R}  + \mu \bar P,
\end{equation}
where $ L^{t} \left(\boldsymbol{q}^{t},\boldsymbol{p}^{t},\boldsymbol{\lambda},\mu \right)={ {{R^{t}} + \sum\nolimits_{k \in \cal K} {{\lambda _k}R_k^{t}}  - \mu {P^{t}}} }$, and
$\boldsymbol{\lambda}=\left[ \lambda_1 , \cdots, \lambda_K \right]$ and $\mu$ are the Lagrangian multipliers.
Then, we have the following dual problem:
\begin{equation}
\mathop {\min}\limits_{\boldsymbol{\lambda}\succcurlyeq 0, \mu \ge 0}\ F\left(\boldsymbol{\lambda},\mu \right), \label{eq.dual}
\end{equation}
where
\begin{equation}
\hspace{-1cm}F\left(\boldsymbol{\lambda},\mu \right) =\mathop {\max}\limits_{ \boldsymbol{q}, \boldsymbol{p} } L\left( {{{\boldsymbol{q}}},{{\boldsymbol{p}}},{\boldsymbol{\lambda }},\mu } \right),\label{eq.L2}\vspace{-15pt}
\end{equation}\addtocounter{equation}{-1}\begin{subequations}
\begin{align}
\hspace{0.7cm}{\rm{s.t.}} \ & 0\le P^{t,f} \le P_{\rm{max}}, \forall f \in \mathcal{F}, \forall s_t  \in \mathcal{S},\label{eq.Pcons}\\
&\boldsymbol{q}^{t} \in Q^{t}, \forall s_t  \in \mathcal{S},
\end{align}
\end{subequations}
%\begin{equation}
% F\left(\boldsymbol{\lambda},\mu \right) =\mathop {\max}\limits_{\scriptstyle0\le P^{t,f} \le P_{\rm{max}}, \forall s_t  \in \mathcal{S},\hfill\atop\scriptstyle \boldsymbol{q}^{t,f} \in Q^{t,f}, \forall s_t  \in \mathcal{S}} L\left( {{{\boldsymbol{q}}},{{\boldsymbol{p}}},{\boldsymbol{\lambda }},\mu } \right),\label{eq.L2}
%\end{equation}
and {$Q^{t}= \left\{\boldsymbol{q}^{t} \left| { 0 \le q_k^{t,f} \le 1,  \forall f \in \mathcal{F}, \forall k \in \mathcal{K}} \ {\text{and}}\ \sum\nolimits_{k \in \mathcal{K}} {q}_k^{t,f} =1, \forall f \in \mathcal{F} \right. \right\}$.
Given Lagrangian multipliers $\boldsymbol{\lambda}$ and $\mu$, we can solve (\ref{eq.L2}) by maximizing $ L^{t} \left(\boldsymbol{q}^{t},\boldsymbol{p}^{t},\boldsymbol{\lambda},\mu \right)$ for each slot $t$:
\begin{equation}
\hspace{-0.8cm}\mathop {\max }\limits_{{\boldsymbol{q}^{t}},\boldsymbol{p}^{t}} \sum\limits_{k \in \cal K} {\left( {1 + {\lambda _k}} \right)R_k^{t}}  - \mu {P^{t}},\vspace{-12pt} \label{eq.P3}\\
\end{equation}\addtocounter{equation}{-1}\begin{subequations}
%\begin{align}
%\hspace{-1.2cm}{\rm{s.t.}} \ & \text{(28a), (28b).}\nonumber
%\end{align}
%\end{subequations}
\begin{align}
\hspace{0cm}{\rm{s.t.}} \ & 0 \le P^{t,f} \le P_{\rm{max}}, \forall f \in \mathcal{F}, \label{eq.Pcons}\\
& {{{\boldsymbol{q}}^{t}} \in {Q^{t}} }. 
\end{align}
\end{subequations}
(\ref{eq.P3}) shows that the optimal $\boldsymbol{q}^{t}$ and $\boldsymbol{p}^{t}$ can be obtained without knowledge of the underlying probability $\pi_t$.
For notational simplicity, we define ${\bar R}^{t} = \sum\limits_{k \in \cal K} {\left( {1 + {\lambda _k}} \right)R_k^{t}} $ which, using (\ref{eq.SR}), can be given by:
\begin{equation}
{\bar R}^{t} = \sum\limits_{f \in \mathcal{F}}  \sum\limits_{k \in \cal K} q_k^{t,f}{\bar R}_k^{t,f} , \label{eq.newRt}
\end{equation}
where
\begin{equation}
{\bar R}_k^{t,f} = \left\{ {\begin{array}{*{20}{l}}
{\left( 1+ \lambda_k \right) \left(\log \left( {1 + \bar \gamma_{{\rm{d}}k}^{t,f}} \right) + \sum\limits_{i \in {\cal R}}  \log \left( {1 + \bar \gamma _{{\rm{a}}i}^{t,f}} \right)\right), {\text{if}}\;k \in {\cal D},}\\
{\left( 1+ \lambda_k \right)\log \left( {1 + \bar \gamma_{{\rm{r}}k}^{t,f}}\right), {\text{if}}\;k \in {\cal R}}.
\end{array}} \right.  \label{eq.newRtk}
\end{equation}
%\begin{equation}
%{\bar R}_k^{t,f} = \left\{ {\begin{array}{*{20}{l}}
%\sum\limits_{k } {q_k^{t,f} {\bar R}_{{\rm{d}}k}^{t,f},\ {\text{if}}\;k \in {\cal D},}\\
%\sum\limits_{k }{q_k^{t,f}{\bar R}_{{\rm{r}}k}^{t,f},\ {\text{if}}\;k \in {\cal R}},
%\end{array}} \right. \label{eq.newRt}
%\end{equation}
%where
%\begin{align}
%{\bar R}_{{\rm{d}}k}^{t,f}&= { \left( 1+ \lambda_k \right) \left(\log \left( {1 + \bar \gamma_{{\rm{d}}k}^{t,f}} \right) + \sum\limits_{i \in {\cal R}}  \log \left( {1 + \bar \gamma _{{\rm{a}}i}^{t,f}} \right)\right),k\in {\cal D}},\label{eq.newRtd}\\
%{\bar R}_{{\rm{r}}k}^{t,f}&={\left( 1+ \lambda_k \right)\log \left( {1 + \bar \gamma_{{\rm{r}}k}^{t,f}}\right),{ k\in {\cal R}}}.\label{eq.newRtr}
%\end{align}
Since (\ref{eq.P3}) is still nonconvex optimization problem, we first determine the optimal $\boldsymbol{p}^{t}$ for given $\boldsymbol{q}^{t}$ and then, obtain the optimal $\boldsymbol{q}^{t}$ under the pre-determined $\boldsymbol{p}^{t}$.
For the fixed $\boldsymbol{p}^{t}$, we can observe from (\ref{eq.newRt}) that $\bar R^{t}$ is an affine function with respect to $q_k^{t,f}$
and thus, the optimal $q_k^{t,f}$ will be one of the boundary conditions (i.e., $q_k^{t,f} \in \left\{0,1\right\}$).
Hence, the gap between the original problem in (\ref{eq.P2}) and the problem after $q_k^{t,f}$ relaxation will be zero.
Given an integer solution $q_k^{t,f}$, we have the following result related to the optimal $\boldsymbol{q}^{t}$ and $\boldsymbol{p}^{t}$.

{\bf{{Proposition 2.}}} The optimal $\boldsymbol{q}^{t}$ and ${\boldsymbol{p}^{t}}$ that solve the optimization problem in (\ref{eq.P3}) are obtained, respectively, as follows:
\begin{equation}
\hat q_k^{t,f} = \left\{ \begin{array}{l}
1, \hspace{5pt} {\text{if }}k = \mathop {\arg \max }\limits_i \left( {{{\left. {\bar R_i^{t,f}} \right|}_{{P^{t,f}} = {{\hat P}_i^{t,f}}}} - \mu {{\hat P}_i^{t,f}}} \right),\\
0,{\text{ otherwise,}}
\end{array} \right.
\hat P^{t,f} = \left\{ \begin{array}{l}
\hat P_k^{t,f},\hspace{5pt} {\text{if }}\hat q_k^{t,f} = 1,\\
0,\hspace{6.5pt} {\text{ otherwise,}}
\end{array} \right.\label{eq.optq}
\end{equation}
for all $k \in \cal K$ and $f \in \mathcal{F}$, where $\hat P_i^{t,f} = \mathop {\arg \max }\limits_{{P^{t,f}}} \left(\bar R_i^{t,f} -  \mu P^{t,f}\right)$ for all $i \in \cal K$ and $f \in \cal F$.
\begin{proof}
The detailed proof is presented in Appendix C.
\end{proof}

\begin{algorithm}[t]
\caption{Resource Allocation Algorithm}
\label{a.2}
\begin{algorithmic}[1]
\State \scriptsize{\textbf{Initialization:} $\boldsymbol{\lambda}^0 =0$, $\mu^0 = 0$, $t =0$, and $f=0$.}
\State \textbf{For each time slot $t$:} $\boldsymbol{\lambda}= \boldsymbol{\lambda}^{t}$ and ${\mu}= \mu^{t}$.
\State \textbf{For each RB $f$:} Calculate $\hat P_i^{t,f} = \mathop {\arg \max }\limits_{{P^{t,f}}}  \left(\bar R_i^{t,f} -  \mu P^{t,f}\right)$, $\forall i \in \cal K$.\\
Calculate $\forall k \in \cal K$,\begin{equation}
\hat q_k^{t,f} = \left\{ \begin{array}{l}
1, \hspace{5pt} {\text{if }}k = \mathop {\arg \max }\limits_i \left( {{{\left. {\bar R_i^{t,f}} \right|}_{{P^{t,f}} = {{\hat P}_i^{t,f}}}} - \mu {{\hat P}_i^{t,f}}} \right),\\
0,{\text{ otherwise,}}
\end{array} \right.\nonumber
\hat P^{t,f} = \left\{ \begin{array}{l}
\hat P_k^{t,f},\hspace{5pt} {\text{if }}\hat q_k^{t,f}  = 1,\\
0,\hspace{6.5pt} {\text{ otherwise.}}
\end{array} \right.\nonumber
\end{equation}\\
Select $\hat k = k$, if $\hat q_k^{t,f}=1$, $\forall k \in \cal{K}.$\\
Phase selection: $\forall k \in \cal R$
\begin{equation}
\boldsymbol{\Phi}_k^{t,f}= \left\{ \begin{array}{l}
\boldsymbol{\hat \Phi}_k^{t,f}, \hspace{5pt} {\text{if }}\hat k \in {\cal R} \hspace{5pt} {\text{and }}\hat q_k^{t,f}=1,\\
\boldsymbol{\bar \Phi}_k^{t,f},\hspace{5pt}{\text{if }}\hat k \in \cal D,\\
-\boldsymbol{I}_N,\hspace{0pt}{\text{ otherwise.}}
\end{array} \right.\nonumber
\end{equation}
\State {Scheduling and power control:} $\boldsymbol{\hat q}^{t,f}\left(  \boldsymbol{\lambda},\mu\right) =\left[ \hat q_1^{t,f}, \cdots, \hat q_K^{t,f} \right]$, $\hat P^{t,f}\left(  \boldsymbol{\lambda},\mu \right) = \hat P^{t,f}$, and $f\leftarrow f + 1$.
\State \textbf{End for}
\State {Update}
${{\boldsymbol{\lambda }}^{ {t + 1} }} = {\left[ {{{\boldsymbol{\lambda }}^{ t }} - {\Delta ^{ t }}{{\boldsymbol{r}}^{ t }}} \right]^ + }$,
${\mu ^{ {t + 1} }} = {\left[ {{\mu ^{ t }} - {\Delta ^{ t }}{{{p}}^{ t }}} \right]^ + }$, and
$t\leftarrow t + 1$.
\State \textbf{End for}
\end{algorithmic}
\end{algorithm}

Proposition 2 shows that, for given $\boldsymbol{\lambda}$ and $\mu$, we can obtain the optimal solution by solving {$\boldsymbol{\hat q}^{t}\left(  \boldsymbol{\lambda},\mu\right) =[ \hat q_k^{t,f}]_{k \in \mathcal{K}, f \in \mathcal{F}}$ and $\boldsymbol{\hat p}^{t}\left(  \boldsymbol{\lambda},\mu\right) = [\hat P^{t,f} ]_{f \in \mathcal{F}}$.
In fact, $\bar \gamma_{{\rm{a}}k}^{t,f}$ and $\bar \gamma_{{\rm{r}}k}^{t,f}$ should be calculated based on Algorithm \ref{a.1} and (\ref{eq.mod}), respectively, to obtain the optimal $\hat P_k^{t,f}$ for all $k \in \cal K$ and $f \in \cal F$,
and this will result in high complexity.
In case of $\bar \gamma_{{\rm{a}}k}^{t,f}$, $\boldsymbol{\bar \Phi}_k^{t,f}$ is used for the proposed PSK modulation and is the pre-determined reflection matrix regardless of its channel state. %, without the need of extensive complexity.
However, in case of $\bar \gamma_{{\rm{r}}k}^{t,f}$, $\boldsymbol{\hat \Phi}_k^{t,f}$ is designed based on Algorithm {\ref{a.1}} and should be updated at each time slot $t$ and RB $f$ for all $k \in \cal K$, resulting in high computational complexity.
Note that Algorithm \ref{a.1} achieves ${\mathcal{O}\left( N^2 \right)}$ in terms of instantaneous SNR regardless of the number of RIS control bits as $N$ increases, as proved in Proposition 1.
Consider an ideal RIS that can achieve an optimal instantaneous SNR using a continuous reflection phases with lossless reflection amplitudes, we can obtain the upper bound of the instantaneous SNR at RUE $k$ as follows: % (i.e., $\left|\Gamma_n \right|=1, \forall n$)
\begin{equation}
 \bar\gamma_{{\rm{r}}k}^{t,f}  \le \hat \gamma_{{\rm{r}}k}^{t,f} %=\frac{{{P^{t,f}}}}{N_0}{ {\left| {\sum\limits_{n = 1}^N {\left| {{f_n^{t,f}}} \right|\left| {{g_{n\hat m}^{t,f}}} \right|} } \right|} ^2}\label{eq.upper}
%\le\hat \gamma_{{\rm{r}}k}^{t,f}  
= \frac{{{P^{t,f}}}}{N_0}{\sum\nolimits_{m = 1}^M {\left| {\sum\nolimits_{n = 1}^N {\left| {{f_n^{t,f}}} \right|\left| {{g_{n,m}^{t,f}}} \right|} } \right|} ^2}.\label{eq.upper}
\end{equation}
%where $ \gamma_k^{t,f} $, $\bar \gamma_k^{t,f}$, and $\hat \gamma_k^{t,f}$ are the instantaneous SNR, the optimal SNR, and their upper bound, respectively,
%and ${\bar{\boldsymbol{\Phi }}_k^{t,f}}$ is the optimal reflection matrix. %, that can be obtained from  \cite{ref.Wu2018beamforming}.
%and $\hat \gamma_k^{t,f}$ is the upper bound of the instantaneous SNR.
In (\ref{eq.upper}), we can observe that $\hat \gamma_{{\rm{r}}k}^{t,f}$ also increases with $\mathcal{O} \left( N^2 \right)$ as $N$ increases, verifying the SNR optimality of Algorithm \ref{a.1}.
Therefore, we can use $\hat \gamma_{{\rm{r}}k}^{t,f}$ instead of $\bar \gamma_{{\rm{r}}k}^{t,f}$ that can significantly reduce the computational complexity.

Next, we solve the dual problem whose goal is to minimize $F\left(\boldsymbol{\lambda},\mu \right)$ in (\ref{eq.dual}).
Since $F\left(\boldsymbol{\lambda},\mu \right)$ is a convex function with respect to $\boldsymbol{\lambda}$ and $\mu$,
we use the stochastic subgradient algorithm as the following iterative updates in \cite{ref.Ermoliev} and \cite{ref.Kall1994stochastic}:
\begin{equation}
{{\boldsymbol{\lambda }}^{ {t + 1} }} = {\left[ {{{\boldsymbol{\lambda }}^{ t }} - {\Delta ^{ t }}{{\boldsymbol{r}}^{ t }}} \right]^ + }, \
{\mu ^{ {t + 1} }} = {\left[ {{\mu ^{ t }} - {\Delta ^{ t }}{{{p}}^{ t }}} \right]^ + },\label{eq.sg1}
\end{equation}
where $\left[ x \right]^+ = \max \left\{ 0,x \right\}$ and $\Delta ^{ t }$ is the step size at time slot.
Also, ${\boldsymbol{r}}^{ t }$ and $p^{t}$ are the stochastic subgradients of $F\left(\boldsymbol{\lambda},\mu \right)$
which can be obtained by using Danskin's theorem in \cite{ref.Bertsekas1999nonlinear}:
\begin{equation}
{{\boldsymbol{r}}^{t}} = \left[ { r_1^{t}, r_2^{t}, \cdots , r_K^{t}} \right], \
{p^{t}} = \bar P - \sum\nolimits_{f \in \cal F} {\hat{P}^{t,f}}\left( {\boldsymbol{\lambda} ^{t},{\mu ^{t}}} \right),
\end{equation}
where $r_k^{t} = {\left. {R_k^{t}} \right|_{\boldsymbol{q}^{t} = \boldsymbol{\hat q}^{t}( \boldsymbol{\lambda}^{t} ,\mu^{t}  ), {\boldsymbol{p}^{t}} = {\hat{\boldsymbol{p}}^{t}}( \boldsymbol{\lambda}^{t} ,\mu^{t}  )}} - \bar R$
and ${\hat{P}^{t,f}}\left( {\boldsymbol{\lambda} ^{t},{\mu ^{t}}} \right) ={\hat{P}^{t,f}}$ for given ${\boldsymbol{\lambda}} ^{t}$ and ${\mu ^{t}}$.}
The stochastic subgradient algorithm in (\ref{eq.sg1}) always converges when we consider the step size of the subgradient algorithm as $\Delta^{t} = 1/t$, as proved in \cite{ref.Ermoliev} and \cite{ref.Kall1994stochastic}.
Since the duality gap goes to zero and the optimality of $\hat \gamma_{{\rm{r}}k}^{t,f}$ is satisfied as $N$ increases to infinity, our algorithm always coverges and the proposed dual optimal solution asymptotically achieves the maximum average sum-rate under the constraints.
The detailed procedure of the proposed algorithm is provided in Algorithm {\ref{a.2}}.

\begin{figure*}[t]\vspace{-2mm}
	\begin{multicols}{2}\vspace{-9mm}
		%\hspace{-4.5mm}
		\includegraphics[width=0.9\columnwidth]{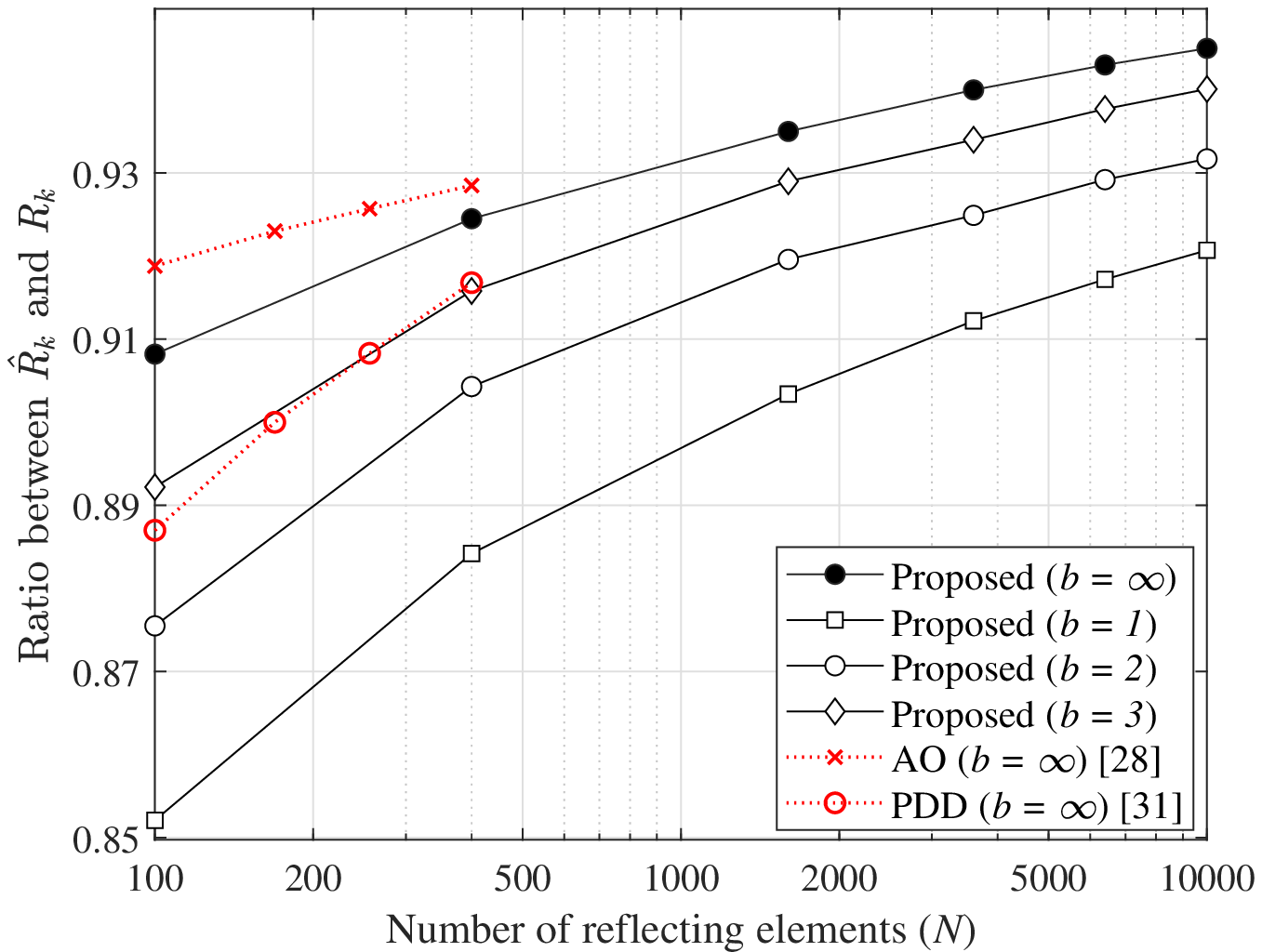}\vspace{-2.5mm}
		\par\caption{\small {Performance comparison of the ergodic rates resulting from Algorithm {\ref{a.1}} with different $b$.}}
\label{fig.10}
		%\hspace{-4.5mm}
		\includegraphics[width=0.9\columnwidth]{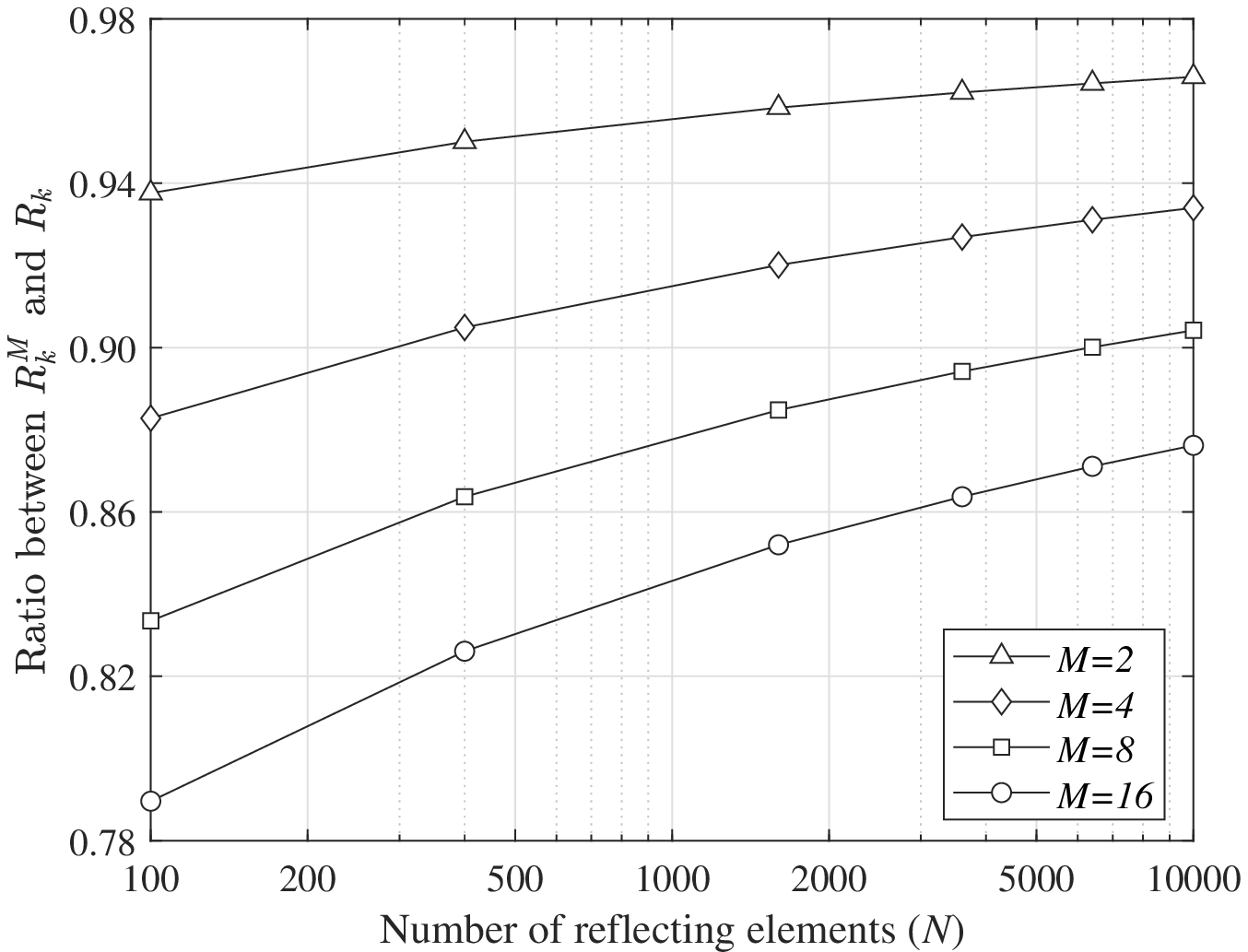}\vspace{-2.5mm}	
		\par\caption{\small {Ergodic rate ratios between Algorithm 1 with MRT and with antenna selection when $b = \infty$.}}
\label{fig.10.1}
		\end{multicols}\vspace{-10mm}
\end{figure*}

\vspace{-0.6cm}
\section{Simulation Results and Analysis}
We run extensive simulations to assess the downlink performance, in terms of the ergodic rate and SER, under a practical-sized RIS environment with finite $N$. %in the considered RIS based MISO system.
We assume that all channels are generated by Rician fading based on (\ref{eq.Gk}), (\ref{eq.fk}), and (\ref{eq.hk}),
where $\kappa_{{\rm{b}}_k} = \kappa_{{\rm{r}}k} = \kappa_{{\rm{d}}k}= 1$.
Given an RIS located in a two-dimensional space perpendicular to the ground, an RIS can be modeled as a uniform planar array (UPA).
Therefore, we use the spatial correlation matrix as used in \cite{ref.Jung2018lisul} in which $\lambda = 0.1$ and the spacing between adjacent reflection elements is set to $\lambda/2$.
We also assume that the DUEs and RUEs are located evenly on a circle centered at the BS with distance of $d_{\rm{bu}}$ and $d_{\rm{br}}$, respectively.
The RISs are also located evenly on a circle centered at the BS with distance of $d_{\rm{br}}$ and the distance between each RIS and RUE is set to $d_{\rm{ru}}$, as shown in Fig. \ref{fig.setup}.
In our simulations, we set $d_{\rm{bu}}=50$ m, $d_{\rm{br}}=100$ m, and $d_{\rm{ru}}=3$ m.
The height of the BS, RIS, and UE are set to $25$ m, $10$ m, and $1.5$ m, respectively,
the path loss exponents for the BS-DUE, BS-RIS, and RIS-RUE links are $3.7$, $2.2$, and $2.2$ with $1$ m reference distance,
and the path loss constant factor is set to $-30$ dB.
Also, the transmission power at the BS and the noise power are set to $-20$ dBm/Hz and $-174$ dBm/Hz, respectively, and $M=2$ unless otherwise stated.
Note that all numerical results are obtained from Monte Carlo simulations that are statistically averaged over a large number of independent runs.

Fig. \ref{fig.10} shows the ratio between the ergodic rate at the RUE, $R_k^{\rm{r}}$, resulting from Algorithm {\ref{a.1}} and theoretical upper bound.
The theoretical upper bound is derived from (\ref{eq.upper}) as follows:
\begin{equation}
\hat R_k = {\rm{E}} \left[ \log \left(1+  \frac{{{P}}}{N_0}{\sum\limits_{m = 1}^M {\left( {\sum\limits_{n = 1}^N {\left| {{f_n^k}} \right|\left| {{g_{n,m}^k}} \right|} } \right)} ^2} \right)\right]. \label{eq.upp_fin}
\end{equation}
We compare the results with the AO and PDD algorithms proposed, respectively, in \cite{ref.Abey2019intelligent} and \cite{ref.MMZ}.
These baselines are shown only up to $N = 400$
due to their computational complexity.
As shown in Fig. \ref{fig.10}, the ergodic rate ratios resulting from the proposed scheme increase toward 1 as $N$ increases,
verifying the optimality of Algorithm {\ref{a.1}} as proved in Proposition 1.
The ergodic rate ratio resulting from the proposed scheme with $b=\infty$ is higher than the one resulting from the PDD algorithm,
since the PDD algorithm does not consider the reflection power loss.
Although the AO algorithm can achieve the upper bound performance, asymptotically,
it requires very high complexity. %.The AO algorithm is complex iterative algorithm 
In both the AO and PDD algorithms, a complexity
in the order of $\mathcal{O}\left(N^2\right)$ is required for each iteration \cite{ref.Abey2019intelligent}, \cite{ref.MMZ}.
However, the proposed Algorithm {\ref{a.1}} requires $\mathcal{O} \left( N \right)$ resulting in a much simpler operation at the BS especially for a large $N$.
Moreover, the AO algorithm uses MRT which requires multiple RF chains resulting in higher cost and hardware complexity compared to the proposed algorithm.
%Furthermore, we can observe from Fig. \ref{fig.10} that Algorithm {\ref{a.1} with $b = 3$ can achieve its upper bound (i.e. the performance of Algorithm {\ref{a.1} with $b = \infty$).
%the upper bound of the proposed Algorithm {\ref{a.1}} (i.e., Algorithm {\ref{a.1} with $b=\infty$) can be achieved when we use Algorithm {\ref{a.1} with $b = 3$.

\begin{figure*}[t]\vspace{-2mm}
	\begin{multicols}{2}\vspace{-9mm}
		%\hspace{-4.5mm}
		\includegraphics[width=0.9\columnwidth]{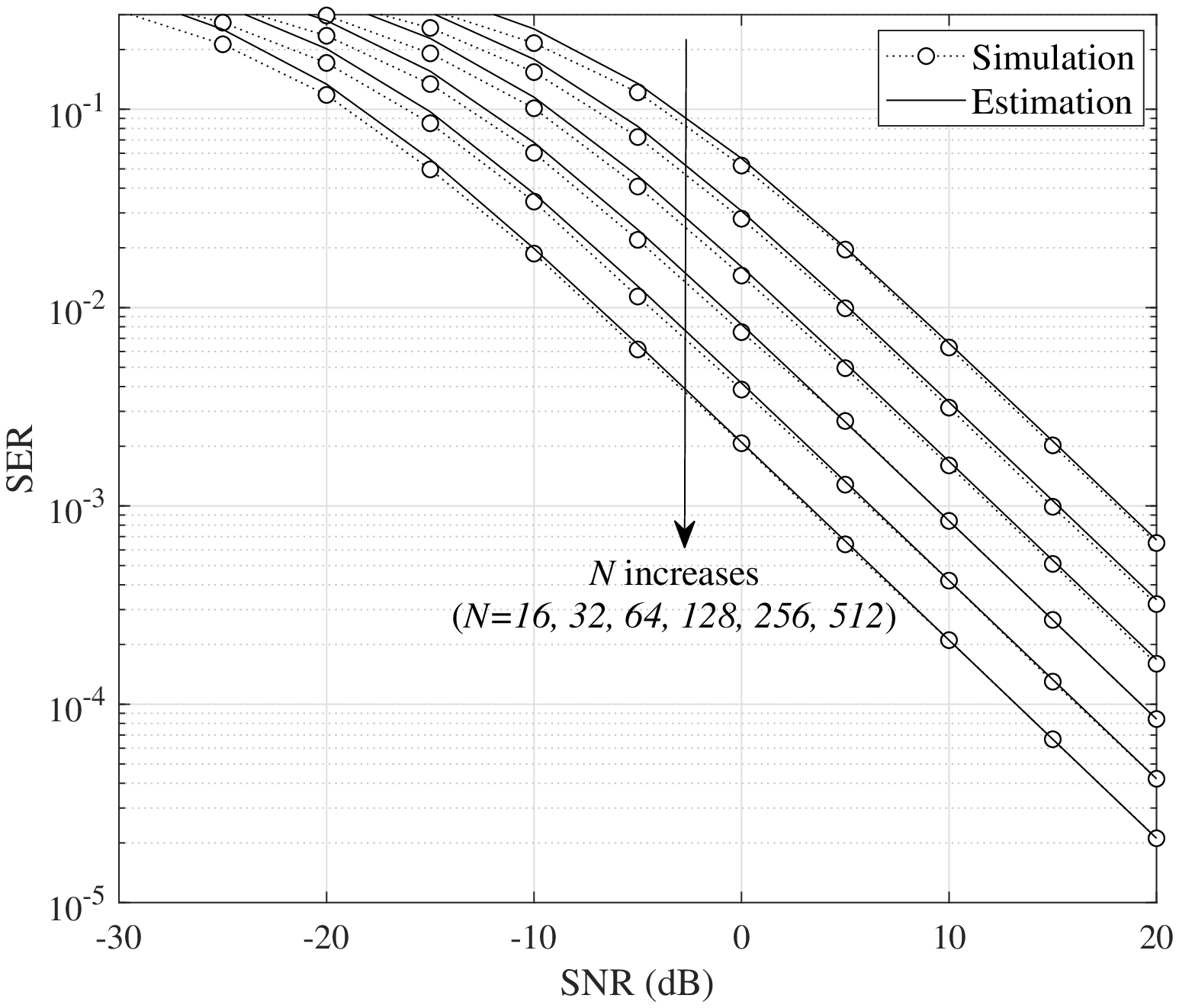}\vspace{-2.5mm}
		\par\caption{\small {Comparison of the average SERs resulting from the proposed modulation with different $N$ values.}}
\label{fig.11}
		%\hspace{-4.5mm}
		\includegraphics[width=1.035\columnwidth]{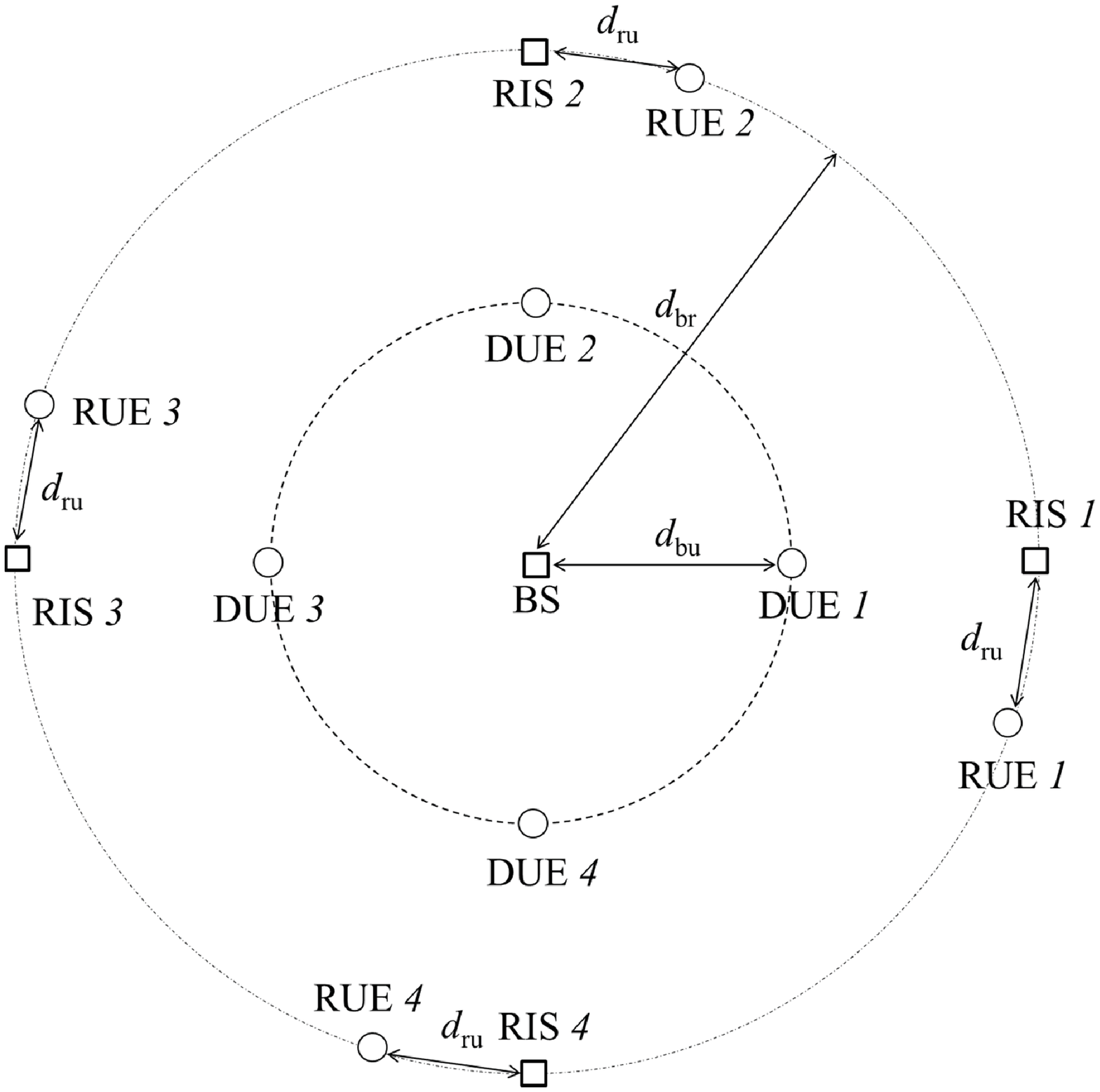}\vspace{-2.5mm}	
		\par\caption{\small {An example of our simulation environment (top view) when $K=8$.}}
\label{fig.setup}
		\end{multicols}\vspace{-10mm}
\end{figure*}

Fig. \ref{fig.10.1} shows the ratios between the ergodic rates resulting from Algorithm 1 with MRT, $R_k^{\rm{M}}$, and with antenna selection for different $M$ values.
As shown in Fig. \ref{fig.10.1}, the ergodic rate ratios decrease as $M$ increases, however, all results increase toward 1 as $N$ increases,
verifying that the transmit antenna selection asymptotically achieves the performance of the MRT.

In Fig. \ref{fig.11}, Theorem 1 is verified in the following scenario.
The BS transmits BPSK signals to the sDUE via a wireless channel, $\boldsymbol{h}_k$,
and also sends the RIS control signals related to the data symbols for the uRIS via a dedicated RIS control link.
Data symbols for the uRIS are modulated based on the proposed modulation technique assuming the BPSK signaling (i.e., $M_0 = 2$) and all channels are generated by independent Rayleigh fading to verify Theorem 1.
As shown in Fig. {\ref{fig.11}}, the asymptotic SERs derived from Theorem 1 are close to the results of our simulations.
Moreover, the SERs linearly decrease as $N$ increases given that the SNR difference is always equal to $3$ dB when $N$ is doubled.
For instance, when the target SER is $2 \cdot 10^{-2}$, the corresponding SNRs are $5$, $2$, $ -1$, $ -4$, $ -7$, $ -10$ dB for $N=16$, $32$, $64$, $128$, $256$, $512$, respectively.
This result shows that the SER can be reduced by increasing $N$ and eventually it converges to zero as $N \to \infty$.

In Figs. \ref{fig.12}--\ref{fig.14}, Algorithm \ref{a.2} is verified in the following scenario.
%In accordance with 3GPP LTE specification \cite{ref.LTE2017TR36211}, we consider a TDMA system in which the system bandwidth is $10$ MHz and consists of $600$ subcarriers each of which equipped with $15$ kHz subcarrier spacing.
%Also, a RB consist of $12$ subcarriers resulting in $50$ RBs within $10$ MHz and $N_{\rm{s}}=6$.
The average transmission power at the BS and the noise power are set to $-20$ dBm/Hz and $-174$ dBm/Hz, respectively,
and 10 UEs (5 DUE and 5 RUE) are located around the BS with $b=1$ unless otherwise stated.
%For experimental simplicity, we consider fixed locations for the RISs and UEs such that the distances from the BS to DUEs, from the BS to RISs, and from the RISs to their serving RUEs are set equally to $5$ m,
%and the distances between the interfering RISs and RUEs are set equally to $10$ m.
We consider an OFDMA system where the system bandwidth is $10$ MHz and the whole bandwidth is equally divided into $25$ RBs (i.e., $F=25$),
and the minimum rate requirement is set to $20$ Mbps per UE.
%and we consider a path loss model as $11 + 20 \log_{10} d\left[ {\rm{m}} \right]$ where $d\left[ {\rm{m}} \right]$ is distance in meters \cite{ref.Tse2005fundamentals}.
%We also assume that $N_{\rm{s}}=6$ and ${{\alpha _i^{t,f}}}=10^{-10}$ for all $t$ and $\forall i \in \cal{D}$.

%\begin{figure}[!ht]
%\centering
%\includegraphics[width=0.5\columnwidth] {figs/fig_IR_bar.eps}
%\caption{Average individual rates at each UE with and without the minimum rate requirements when $N=100$ and $b=1$.}
%\label{fig.13}
%\end{figure}
%
%\begin{figure}[!ht]
%\centering
%\includegraphics[width=0.5\columnwidth] {figs/fig_SR_N.eps}
%\caption{Performance comparison of the average sum-rates for cases with and without the minimum rate requirements.}
%\label{fig.14}
%\end{figure}

\begin{figure}[t]
\centering
\includegraphics[width=0.42\columnwidth] {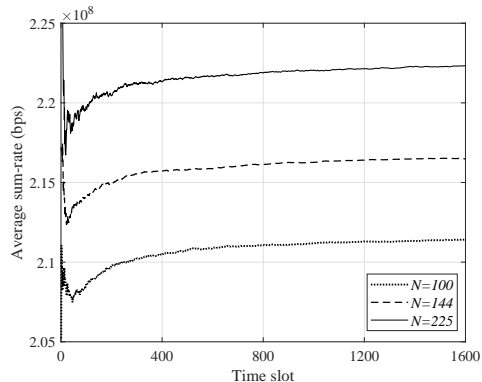}\vspace{-0.5cm}
\caption{Convergence of the average sum-rate resulting from Algorithm \ref{a.2}.}\vspace{-0.8cm}
\label{fig.12}
\end{figure}

Fig. \ref{fig.12} shows the convergence of the average sum-rate resulting from Algorithm {\ref{a.2}}.
This convergence shows that Algorithm {\ref{a.2}} always satisfies the constraints of the transmission power and the rate requirements in (\ref{eq.P2}) regardless of $N$,
verifying the convergence of Algorithm {\ref{a.2}}.
Also, the convergence is satisfied within seconds after initial access (e.g., $1$ ms $\times$$1000$ TTIs = $1.0$ s) regardless of $N$,
showing that the optimal sum-rate can be achieved within a few seconds.

Figs. \ref{fig.13} and \ref{fig.14} compare the average data rates resulting from Algorithm {\ref{a.2}} with and without the minimum rate requirements.
As proved in Proposition 1, the received SNRs at the RUEs increase with $\mathcal{O} \left(N^2 \right)$ while those at the DUEs approximately keep constant as $N$ increases.
For a large $N$ without considering the rate requirements, Algorithm {\ref{a.2}} selects more RUEs than DUEs for better sum-rate resulting in unfairness of individual rates, as shown in Fig. \ref{fig.13}.
However, in this case, we can achieve better performance in terms of the average sum-rate compared to the case in which the UEs have the requirements on
the minimum rates, as shown in Fig. \ref{fig.14}.
On the other hand, when we consider the minimum rate requirements, all individual rates satisfy those requirements
while the average sum-rate decreases, as shown in Figs. \ref{fig.13} and \ref{fig.14}.
Hence, Algorithm {\ref{a.2}} can control the tradeoff between fairness and maximum performance. % according to the system requirements.
Since the BS serves a single UE at each RB, 
the user selection gain resulting from $\hat R = \mathop {\max}\limits_{k \in \mathcal{R}} \hat R_k$ will be limited at high SNR region. % even for large $K$. %resulting from a large $N$.
%Therefore, as shown in Fig. \ref{fig.15}, the upper bounds for different $K$ (e.g., $K=20,40,100$) converge equally to the same value as $N$ increases.
However, since the additional sum-rate is calculated by the sum of $\left| \mathcal{R} \right|$ additional rates from all uRUEs, the additional sum-rate linearly increases as $K$ increases. %as shown in Fig. \ref{fig.15}.
This additional sum-rate gradually increases its effects on the average sum-rate as $N$ and $K$ increase as shown in Fig. \ref{fig.14},
showing that our algorithm can support massive connectivity.

\begin{figure*}[t]\vspace{-2mm}
	\begin{multicols}{2}\vspace{-9mm}
		%\hspace{-4.5mm}
		\includegraphics[width=0.9\columnwidth]{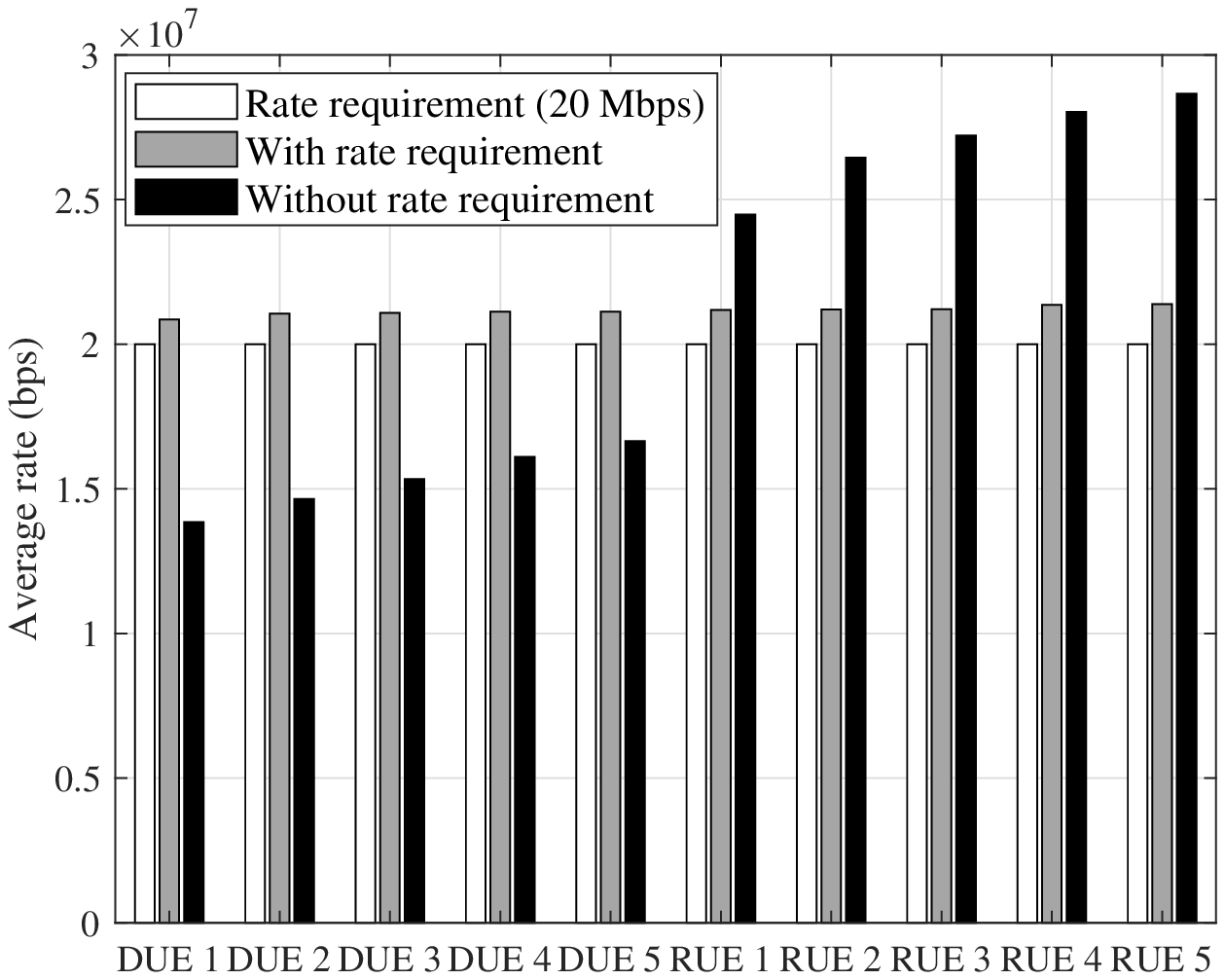}\vspace{-2.5mm}
		\par\caption{\small Average individual rates at each UE with and without $\bar R$ when $N=100$.}
\label{fig.13}
		%\hspace{-4.5mm}
		\includegraphics[width=0.9\columnwidth]{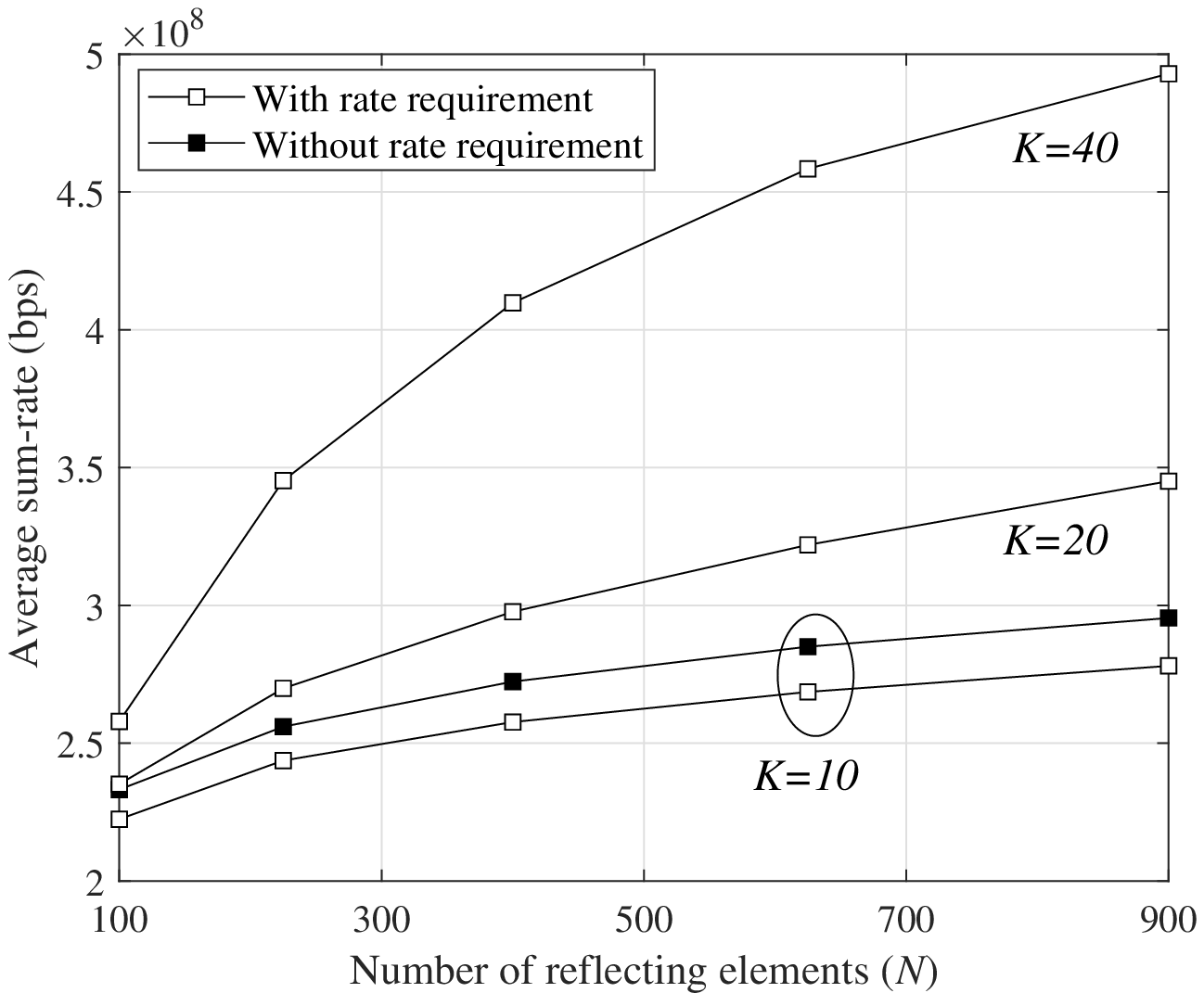}\vspace{-2.5mm}	
		\par\caption{\small {Performance comparison of the average sum-rates for cases with and without $\bar R$.}}
\label{fig.14}
		\end{multicols}\vspace{-10mm}
\end{figure*}

\vspace{-12pt}
\section{Conclusions}
In this paper, we have asymptotically analyzed the optimality of the achievable rate using practical RISs in presence of limitations such as practical reflection coefficients and limited RIS control link capacity.
In particular, we have designed a passive beamformer that can achieve the asymptotic optimal SNR under discrete reflection phases with a practical reflection power loss, %resulting from practical reflection coefficients,
and shown that it achieves a SNR optimality even with one bit RIS control.
%We have also proposed a modulation scheme that can be used in a downlink RIS system resulting in higher sum-rate and proved that it achieves an asymptotic SNR of conventional massive array systems such as massive MIMO and MIMO relays.
%Moreover, we have derived the approximated SER of the proposed modulation technique, showing that the asymptotic SNR of the proposed modulation follows the chi-squared distribution regardless of $M$.
We have also proposed a modulation scheme that can be used in a downlink RIS system resulting in higher achievable sum-rate than a conventional network without RIS.
Moreover, we have derived the approximated SER of the proposed modulation scheme, showing that it achieves an asymptotic SNR of a conventional massive array systems such as a massive MIMO or MIMO relay system.
Furthermore, we have proposed the resource allocation algorithm under consideration of the aforementioned passive beamforming and the modulation schemes that achieves the asymptotic optimal sum-rate.
We have shown that the proposed algorithms can analytically achieve the performance of an ideal RIS.
Simulation results have shown that the results of our algorithms converge to the asymptotic upper bound as $N \to \infty$.
In particular, we have observed that the approximated SER is in close agreement with the result from our simulations
and the proposed resource allocation algorithm asymptotically achieves the optimal performance satisfying the minimum rate requirements at each UE.
Moreover, our results have shown that the proposed resource algorithm can control the tradeoff between fairness of individual performance and maximum system performance. %, according to the system requirements.
We finally have shown that the performance of our algorithm increases as the number of UEs increases, resulting from the additional rate achieved at the uRUEs.
Therefore, we expect that our algorithm will be invaluable solution for future wireless networks supporting massive connectivity.
Our future work will include extending our results
to additional practical scenarios such as a multi-user MIMO OFDM system 
that has an RIS environment with a finite number of reflecting elements and multi-hop reflections.

\vspace{-0.4cm}
\section*{Appendix A \\ Proof of Lemma 1}\vspace{-0.2cm}
Given that $\left| f_n^k\right|$ and $\left| g_{n,m_0}^k\right|$ are independent random variables that follow a non-central chi distribution with two degrees of freedom,
the mean and variance of $\munderbar \gamma_{\rm{l}}^k$ are obtained, respectively, as follows: ${\rm{E}}\big[ {{{\munderbar \gamma }_{\rm{l}}^k}} \big] = \sum\nolimits_{n=1}^{N}\bar \mu _{k,n}^{\rm{b}} \bar\mu _{k,n}^{\rm{r}}$,
\begin{equation}
{\rm{Var}}\big[ {{{\munderbar \gamma }_{\rm{l}}^k}} \big] =\sum\limits_{n = 1}^N {\left( {{{\left( {\bar\sigma _{k,n}^{\rm{b}}\bar\sigma _{k,n}^{\rm{r}}} \right)}^2} + {{\left( {\bar\mu _{k,n}^{\rm{b}}\bar\sigma _{k,n}^{\rm{r}}} \right)}^2} + {{\left( {\bar\mu _{k,n}^{\rm{r}}\bar\sigma _{k,n}^{\rm{b}}} \right)}^2}} \right)} 
+  \sum\limits_{n\ne j} {\rm{Cov}}\big[\left| f_n^k \right|\left| g_{n,m_0}^k\right|, \left| f_j^k\right|\left| g_{j,m_0}^k\right| \big],\nonumber
\end{equation}
where
\begin{align}
\bar\mu _{k,n}^{\rm{b}} &= \sqrt {\frac{\pi }{4  \left({{{{\kappa_{{\rm{b}}k}} + 1}}}\right)}\sum\nolimits_{j = 1}^N {\left| {{{\left( {{{\boldsymbol{R}}_{{\rm{b}}k}}} \right)}_{n,j}}} \right|} } {L_{{{\frac{1}{ 2}}}}}\left( { - \frac{{{{{\kappa_{{\rm{b}}k}}\left| {\bar g_{n,m_0}^k} \right|}^2}}}{{\sum\nolimits_{j = 1}^N {\left| {{{\left( {{{\boldsymbol{R}}_{{\rm{b}}k}}} \right)}_{n,j}}} \right|} }}} \right),\\
\bar\mu _{k,n}^{\rm{r}} &= \sqrt {\frac{\pi }{4\left({\kappa_{{\rm{r}}k}}+1\right)}\sum\nolimits_{j = 1}^N {\left| {{{\left( {{{\boldsymbol{R}}_{{\rm{r}}k}}} \right)}_{n,j}}} \right|} } {L_{{{\frac{1}{ 2}}}}}\left( { - \frac{{{{{\kappa_{{\rm{r}}k}}\left| {\bar f_n^k} \right|}^2}}}{{\sum\nolimits_{j = 1}^N {\left| {{{\left( {{{\boldsymbol{R}}_{{\rm{r}}k}}} \right)}_{n,j}}} \right|} }}} \right),
\end{align}
\vspace{-0.2cm}
\begin{align}
{{\bar\sigma _{k,n}^{\rm{b}}}} &= \sqrt{\frac{\sum\nolimits_{j} {\big| {{{\left( {{{\boldsymbol{R}}_{{\rm{b}}k}}} \right)}_{n,j}}} \big|} + {\kappa_{{\rm{b}}k}}{\left| {\bar g_{n,m_0}^k} \right|^2}}{{\kappa_{{\rm{b}}k}}+1}  - {\left( {\bar\mu _{k,n}^{\rm{b}}} \right)^2}},
{ {\bar\sigma _{k,n}^{\rm{r}}} } = \sqrt{\frac{\sum\nolimits_{j} {\big| {{{\left( {{{\boldsymbol{R}}_{{\rm{r}}k}}} \right)}_{n,j}}} \big|}+ {\kappa_{{\rm{r}}k}}{\left| {\bar f_n^k} \right|^2} }{{\kappa_{{\rm{r}}k}}+1}  - {\left( {\bar\mu _{k,n}^{\rm{r}}} \right)^2}}.\nonumber
\end{align}
Since ${\rm{Cov}}\big[\left| f_n^k \right|\left| g_{n,m_0}^k\right|, \left| f_j^k\right|\left| g_{j,m_0}^k\right| \big] \ge 0$ for all $n$ and $j$, ${\rm{E}}\big[|\munderbar \gamma_{\rm{l}}^k|^2\big]$ is lower bounded by
\begin{equation}
{\rm{E}}\big[|\munderbar \gamma_{\rm{l}}^k|^2\big] \ge \left(\sum\nolimits_{n=1}^{N} \bar\mu _{k,n}^{\rm{b}} \bar\mu _{k,n}^{\rm{r}} \right)^2 + \sum\nolimits_{n = 1}^N {\left( {{{\left( {\bar\sigma _{k,n}^{\rm{b}}\bar\sigma _{k,n}^{\rm{r}}} \right)}^2} + {{\left( {\bar\mu _{k,n}^{\rm{b}}\bar\sigma _{k,n}^{\rm{r}}} \right)}^2} + {{\left( {\bar\mu _{k,n}^{\rm{r}}\bar\sigma _{k,n}^{\rm{b}}} \right)}^2}} \right)},\label{eq.A.SNRl}
\end{equation}
where $\bar\mu _{k,n}^{\rm{b}}> 0$ and $\bar\mu _{k,n}^{\rm{r}} >0$ since $L_{n}\left(x\right) >0$ for $x<0$.
Similarly, ${\rm{E}}\big[|\munderbar \gamma_{{\rm{r}}m}^k|^2\big]$ is bounded by
\begin{equation}
{\rm{E}}\big[|\munderbar \gamma_{{\rm{r}}m}^k|^2\big] \ge \Big|\sum\limits_{n} \tilde\mu _{k,n,m}^{\rm{b}} \left(\tilde\mu _{k,n}^{\rm{r}}\right)^* \Big|^2 + \sum\limits_{n} {\left( {{{\left( {\tilde\sigma _{k,n}^{\rm{b}}\tilde\sigma _{k,n}^{\rm{r}}} \right)}^2} + {{\left( {\left|\tilde\mu _{k,n,m}^{\rm{b}}\right|\tilde\sigma _{k,n}^{\rm{r}}} \right)}^2} + {{\left( {\left|\tilde\mu _{k,n}^{\rm{r}}\right|\tilde\sigma _{k,n}^{\rm{b}}} \right)}^2}} \right)},\label{eq.A.SNRr}
\end{equation}
where
\begin{equation}
\tilde\mu _{k,n,m}^{\rm{b}} {=} \sqrt{\frac{{\kappa_{{\rm{b}}k}}}{{\kappa_{{\rm{b}}k}+1}}}{{\bar g_{n,m}^k} },
\tilde\mu _{k,n}^{\rm{r}} {=} \sqrt{\frac{{\kappa_{{\rm{r}}k}}}{{\kappa_{{\rm{r}}k}+1}}}{{\bar f_{n}^k} },
\tilde\sigma _{k,n}^{\rm{b}} {=}\sqrt{\frac{\sum\nolimits_{j = 1}^N {\big| {{{\left( {{{\boldsymbol{R}}_{{\rm{b}}k}}} \right)}_{n,j}}} \big|}}{{\kappa_{{\rm{b}}k}+1}}},
\tilde\sigma _{k,n}^{\rm{r}} {=} \sqrt{\frac{\sum\nolimits_{j = 1}^N {\big| {{{\left( {{{\boldsymbol{R}}_{{\rm{r}}k}}} \right)}_{n,j}}} \big|}}{{\kappa_{{\rm{r}}k}+1}}}.\nonumber %\label{eq.fin}
\end{equation}
%\begin{gather}
% \tilde\mu _{k,i,m}^{\rm{b}} = \sqrt{\frac{{\kappa_{{\rm{b}}k}}}{{\kappa_{{\rm{b}}k}+1}}}{{\bar g_{i,m}^k} },\\
% \tilde\mu _{k,i}^{\rm{r}} = \sqrt{\frac{{\kappa_{{\rm{r}}k}}}{{\kappa_{{\rm{r}}k}+1}}}{{\bar f_{i}^k} },\\
% \tilde\sigma _{k,i}^{\rm{b}} = \sqrt{\frac{\sum\nolimits_{j = 1}^N {\left| {{{\left( {{{\boldsymbol{R}}_{{\rm{b}}k}}} \right)}_{i,j}}} \right|}}{{\kappa_{{\rm{b}}k}+1}}},\\
% \tilde\sigma _{k,i}^{\rm{r}} = \sqrt{\frac{\sum\nolimits_{j = 1}^N {\left| {{{\left( {{{\boldsymbol{R}}_{{\rm{r}}k}}} \right)}_{i,j}}} \right|}}{{\kappa_{{\rm{r}}k}+1}}}.
%\end{gather}
Given the definition of $ {{{\munderbar \gamma }_k}}$ from (\ref{eq.SNRlb}), we have
\begin{align}
&E\big[ {{{\munderbar \gamma }_k}} \big] \ge \frac{{{PE_{{k}}^{\rm{r}}}}\left| \Gamma \right|_{\rm{min}}^2}{N_0}\left({\rm{E}}\big[|\munderbar \gamma_{\rm{l}}^k|^2\big] + \sum\nolimits_{m \ne m_0}^M{\rm{E}}\big[|\munderbar \gamma_{{\rm{r}}m}^k|^2\big] \right)\nonumber\\
&\ge \frac{{{PE_{{k}}^{\rm{r}}}}\left| \Gamma \right|_{\rm{min}}^2}{N_0}\left\{ \left(\sum\nolimits_{n=1}^{N} \bar\mu _{k,n}^{\rm{b}} \bar\mu _{k,n}^{\rm{r}} \right)^2 + \sum\nolimits_{n = 1}^N {\left( {{{\left( {\bar\sigma _{k,n}^{\rm{b}}\bar\sigma _{k,n}^{\rm{r}}} \right)}^2} + {{\left( {\bar\mu _{k,n}^{\rm{b}}\bar\sigma _{k,n}^{\rm{r}}} \right)}^2} + {{\left( {\bar\mu _{k,n}^{\rm{r}}\bar\sigma _{k,n}^{\rm{b}}} \right)}^2}} \right)}\right. \nonumber\\
&+ \left. \sum\limits_{m \ne m_0}^M \left(\left|\sum\limits_{n=1}^{N} \tilde\mu _{k,n,m}^{\rm{b}} \tilde\mu _{k,n}^{{\rm{r}}*} \right|^2 + \sum\limits_{n = 1}^N {\left( {{{\left( {\tilde\sigma _{k,n}^{\rm{b}}\tilde\sigma _{k,n}^{\rm{r}}} \right)}^2} + {{\left( {\left|\tilde\mu _{k,n,m}^{\rm{b}}\right|\tilde\sigma _{k,n}^{\rm{r}}} \right)}^2} + {{\left( {\left|\tilde\mu _{k,n}^{\rm{r}}\right|\tilde\sigma _{k,n}^{\rm{b}}} \right)}^2}} \right)} \right)\right\}\label{eq.A.SNRlb}
\end{align} %{\left( {\sum\limits_{i = 1}^N {\bar\mu _{k,i}^{\rm{b}}\bar\mu _{k,i}^{\rm{r}}} } \right)^2}
In order to verify the scaling law of (\ref{eq.A.SNRlb}), we first determine the scaling law of $\left(\sum\nolimits_{n} \bar\mu _{k,n}^{\rm{b}} \bar\mu _{k,n}^{\rm{r}} \right)^2$ in (\ref{eq.A.SNRlb}) according to $N$.
Since we consider an RIS located in a two-dimensional space perpendicular to the ground, it can be modeled as a UPA \cite{ref.Jung2018lisul,ref.Song2017UPA}.
Hence, we have $\sum\nolimits_{j = 1}^N {| {{{\left( {{{\boldsymbol{R}}_{{\rm{b}}k}}} \right)}_{n,j}}} |}=\sum\nolimits_{j = 1}^N {| {{{\left( {{{\boldsymbol{R}}_{{\rm{b}}k}}} \right)}_{n,j}}} |}=1$.
From the scaling law for $N$,  $\left(\sum\nolimits_{n} \bar\mu _{k,n}^{\rm{b}} \bar\mu _{k,n}^{\rm{r}} \right)^2$ in (\ref{eq.A.SNRlb}) is then calculated by the squared sum of positive $N$ elements and increases with $\mathcal{O} \left( N^2 \right)$.
On the other hand, $\sum\nolimits_{n} {\left( {{{\left( {\bar\sigma _{k,n}^{\rm{b}}\bar\sigma _{k,n}^{\rm{r}}} \right)}^2} + {{\left( {\bar\mu _{k,n}^{\rm{b}}\bar\sigma _{k,n}^{\rm{r}}} \right)}^2} + {{\left( {\bar\mu _{k,n}^{\rm{r}}\bar\sigma _{k,n}^{\rm{b}}} \right)}^2}} \right)}$ in (\ref{eq.A.SNRlb}) is calculated by the sum of positive $N$ elements
resulting in $\mathcal{O} \left( N \right)$ and becomes negligible compared to $\left(\sum\nolimits_{n} \bar\mu _{k,n}^{\rm{b}} \bar\mu _{k,n}^{\rm{r}} \right)^2$ as $N$ increases.
Similarly, the other term in (\ref{eq.A.SNRlb}) follows $\mathcal{O} \left( N \right)$ and also become negligible as $N$ increases.
Therefore, (\ref{eq.A.SNRlb}) eventually converges to $\frac{{{PE_{{k}}^{\rm{r}}}}\left| \Gamma \right|_{\rm{min}}^2}{N_0}{\left( {\sum\nolimits_{n} {\bar\mu _{k,n}^{\rm{b}}\bar\mu _{k,n}^{\rm{r}}} } \right)^2}$, which completes the proof.
%\begin{equation}
%E\big[ {{{\munderbar \gamma }_k}} \big] \ge \frac{{{PE_{{k}}^{\rm{r}}}}\left| \Gamma \right|_{\rm{min}}^2}{N_0}{\left( {\sum\limits_{i = 1}^N {\bar\mu _{k,i}^{\rm{b}}\bar\mu _{k,i}^{\rm{r}}} } \right)^2},
%\end{equation}
%$E\big[ {{{\munderbar \gamma }_k}} \big]$ increases with $\mathcal{O} \left( N^2 \right)$

\vspace{-0.2cm}
\section*{Appendix B \\ Proof of Theorem 1}
Given that the maximum instantaneous SNR at sDUE $k$ can be achieved by using the MRT such as ${\boldsymbol{w}_k}={\boldsymbol{h}_k}/\left\|{\boldsymbol{h}_k}\right\|$,
the instantaneous SNR at uRUE $i$ can be derived by using (\ref{eq.yi2}) as
\begin{equation}
\gamma _i^{{\rm{uR}}} = \frac{{PA{{\left( {{\omega _i}} \right)}^2}{N_{\rm{s}}}E_k^{\rm{d}}{{\left| {{\boldsymbol{f}}_i^{\rm{H}}{{\boldsymbol{G}}_i}{{\boldsymbol{h}}_k}} \right|}^2}}}{{{N_0}{{\left\| {{{\boldsymbol{h}}_k}} \right\|}^2}}}. \label{eq.uSNR}
\end{equation}
Since the uRIS transmits the same symbols during the data transmission period, the uRUE achieves the diversity gain proportional to $N_{\rm{s}}$ resulting in $N_{\rm{s}}$-fold of the desired signal power as shown in (\ref{eq.uSNR}).
In order to verify the scaling law of the average SNR at uRUE $i$, we first determine the scaling law of ${\rm{E}}\left[{{| {{\boldsymbol{f}}_i^{\rm{H}}{{\boldsymbol{G}}_i}{{\boldsymbol{w}}_k}} |}^2}\right]$ for $N$.
From (\ref{eq.A.SNRr}), ${\rm{E}}\left[{{| {{\boldsymbol{f}}_i^{\rm{H}}{{\boldsymbol{G}}_i}{{\boldsymbol{w}}_k}} |}^2}\right]$ is bounded as
\begin{equation}
{\rm{E}}\left[{{| {{\boldsymbol{f}}_i^{\rm{H}}{{\boldsymbol{G}}_i}{{\boldsymbol{w}}_k}} |}^2}\right] \ge \left|\sum\nolimits_{m}{{\rm{E}}\left[w^k_m\right]}\munderbar \mu_{i,m}\right|^2 + \sum\nolimits_{m}\left({\rm{Var}} {\left[w^k_m\right]}\munderbar \sigma_{i,m}^2
+ \left|\munderbar \mu_{i,m}\right|^2{\rm{Var}}{\left[w^k_m\right]} + \left|{\rm{E}} {\left[w^k_m\right]}\right|^2\munderbar \sigma_{i,m}^2\right),\nonumber
\end{equation}
where
$\munderbar \mu_{i,m}{=}   \sum\nolimits_{n=1}^{N} \tilde\mu _{i,n,m}^{\rm{b}} \left(\tilde\mu _{i,n}^{\rm{r}}\right)^*$ and
$\munderbar \sigma_{i,m}^2{=} \sum\nolimits_{n = 1}^N {\left( {{{\left( {\tilde\sigma _{i,n}^{\rm{b}}\tilde\sigma _{i,n}^{\rm{r}}} \right)}^2} + {{\left( {\left|\tilde\mu _{i,n,m}^{\rm{b}}\right|\tilde\sigma _{i,n}^{\rm{r}}} \right)}^2} + {{\left( {\left|\tilde\mu _{i,n}^{\rm{r}}\right|\tilde\sigma _{i,n}^{\rm{b}}} \right)}^2}} \right)}$.
%Here, $\tilde\mu _{i,n,m}^{\rm{b}}$, $\tilde\mu _{i,n}^{\rm{r}}$, $\tilde\sigma _{i,n}^{\rm{b}}$, and $\tilde\sigma _{i,n}^{\rm{r}}$ are given in (\ref{eq.fin}).
From the scaling law for $N$, the lower bound increases with $\mathcal{O} \left( N \right)$ as $N$ increases.
Therefore, the uRUE can achieve average SNR in order of  $\mathcal{O} \left( N \right)$.

As a special case, we can analyze the average SNR at uRUE $i$ for i.i.d. Rayleigh fading channels by assuming that ${\kappa_{{\rm{b}}k}}={\kappa_{{\rm{r}}k}}={\kappa_{{\rm{d}}k}}=0$ and ${\boldsymbol{R}_{{\rm{b}}k}}={\boldsymbol{R}_{{\rm{r}}k}}={\boldsymbol{I}}_N$.
In order to analyze the impact on (\ref{eq.uSNR}) of the number of BS antennas, we consider two exteme cases: $M=1$ and $M \to \infty$.
We first analyze (\ref{eq.uSNR}) when $M=1$.
Then, (\ref{eq.uSNR}) is obtained by
%\begin{equation}
$\gamma _i^{{\rm{uR}}} = {{PA{{\left( {{\omega _i}} \right)}^2}{N_{\rm{s}}}E_k^{\rm{d}}{{\left| {\sum\nolimits_{n = 1}^N {f_n^{i*}{g_n^i}} } \right|}^2}}}/{{{N_0}}}.$ %\label{eq.gamma_ur}
%\end{equation}
By using the CLT, ${\sum\nolimits_{n = 1}^N {f_n^{i*}{g_n^i}} }$ converges to a complex Gaussian distributed random variable with zero mean and variance of $N$. %, as $N$ increases.
Then, we have
\begin{equation}
\frac{1}{N}\gamma _i^{{\rm{uR}}} \xrightarrow[N \to \infty ]{\rm{d}} \frac{{PA{{\left( {{\omega _i}} \right)}^2}{N_{\rm{s}}}E_k^{\rm{d}}}}{{2{N_0}}}\chi _2^2,\label{eq.Chi}
\end{equation}
where `` $\xrightarrow[M \to \infty ]{\rm{d}}$'' denotes the convergence in distribution and
$\chi _k^2$ denotes the chi-squared distribution with $k$ degress of freedom.
We consider a random variable $Y_1$ as $\gamma _i^{{\rm{uR}}}$ when $M=1$.
Then, the mean and the probability density function (PDF) of $Y_1$ can be obtained, respectively, as
\begin{equation}
{\rm{E}}\left[ {{Y_1}} \right] =  \frac{{PA{{\left( {{\omega _i}} \right)}^2}N{N_{\rm{s}}}E_k^{\rm{d}}}}{{{N_0}}},
{f_{{Y _1}}}\left( y  \right) = \frac{{{N_0}}}{{PA{{\left( {{\omega _i}} \right)}^2}N{N_{\rm{s}}}E_k^{\rm{d}}}}{e^{ - \frac{{{N_0}y }}{{PA{{\left( {{\omega _i}} \right)}^2}N{N_{\rm{s}}}E_k^{\rm{d}}}}}}.\label{eq.PDFY1}
\end{equation}
Next, we analyze (\ref{eq.uSNR}) as $M \to \infty$. Then, (\ref{eq.uSNR}) can be obtained by:
\begin{equation}
\gamma _i^{{\rm{uR}}} = \mathop {\lim }\limits_{M \to \infty } \frac{{PA{{\left( {{\omega _i}} \right)}^2}{N_{\rm{s}}}E_k^{\rm{d}}{{\left| {\sum\nolimits_{m = 1}^M {{h_m^k}} \sum\nolimits_{n = 1}^N {f_n^{i*}{g_{n,m}^i}} } \right|}^2}}}{{{N_0}{{\sum\nolimits_{m = 1}^M {\left| {{h_m^k}} \right|} }^2}}},\label{eq.SNRinfM}
\end{equation}
where ${{\bf{h}}_k} = {\left[ {{h_1^k}, \cdots ,{h_M^k}} \right]^{\rm{T}}}$.
On the basis of the CLT for a large $M$, ${\sum\nolimits_{m = 1}^M {{h_m^k}} \sum\nolimits_{n = 1}^N {f_n^{i*}{g_{n,m}^i}} }$ in (\ref{eq.SNRinfM}) converges to a complex Gaussian distributed random variable with zero mean and variance of $NM$.
Hence, we have
%\begin{equation}
$\frac{1}{M}{{{\left| {\sum\nolimits_{m = 1}^M {{h_m^k}} \sum\nolimits_{n = 1}^N {f_n^{i*}{g_{n,m}^i}} } \right|}^2}}  \xrightarrow[M \to \infty ]{\rm{d}} \frac{N}{2}\chi _2^2$.
%\end{equation}
Also, the denominator in (\ref{eq.SNRinfM}) follows the chi-squared distribution: $N_0{{{\sum\nolimits_{m = 1}^M {\left| {{h_m^k}} \right|} }^2}} \sim \frac{N_0}{2}\chi _{2M}^2$.
Note that the distribution of the ratio of two chi-squared distributed random variables is the beta prime distribution \cite{ref.Johnson1995continuous}.
Considering a random variable $Y_M$ as $\gamma _i^{{\rm{uR}}}$ with a large $M$,
the mean and the PDF of $Y_M$ can be obtained based on the beta prime distribution, respectively:
\begin{gather}
{\rm{E}}\left[ {{Y_M}} \right] = \frac{{PA{{\left( {{\omega _i}} \right)}^2}NM{N_{\rm{s}}}E_k^{\rm{d}}}}{{{N_0}\left( {M - 1} \right)}},\label{eq.eYM}\\
{f_{{Y_M}}}\left( y \right) = \frac{{{N_0}}}{{PA{{\left( {{\omega _i}} \right)}^2}N{N_{\rm{s}}}E_k^{\rm{d}}}}{\left( {1 + \frac{{{N_0}y}}{{PA{{\left( {{\omega _i}} \right)}^2}NM{N_{\rm{s}}}E_k^{\rm{d}}}}} \right)^{ - M - 1}.}\label{eq.fYM}
\end{gather}
As $M \to \infty$, (\ref{eq.eYM}) and (\ref{eq.fYM}) are obtained, respectively, as follows:
\begin{align}
{\rm{E}}\left[ {{Y_\infty }} \right] &=  \mathop {\lim }\limits_{M \to \infty } {\rm{E}}\left[ {{Y_M}} \right] ={{PA{{\left( {{\omega _i}} \right)}^2}N{N_{\rm{s}}}E_k^{\rm{d}}}}/{{{N_0}}},\label{eq.EYinf}\\
{f_{{Y_\infty }}}\left( y \right)& = \mathop {\lim }\limits_{M \to \infty } {f_{{Y_M}}}\left( y \right)\mathop  = \limits_{(c)} \frac{{{N_0}}}{{PA{{\left( {{\omega _i}} \right)}^2}N{N_{\rm{s}}}E_k^{\rm{d}}}}{e^{ - \frac{{{N_0}y}}{{PA{{\left( {{\omega _i}} \right)}^2}N{N_{\rm{s}}}E_k^{\rm{d}}}}}},\label{eq.PDFYinf}
\end{align}
where $(c)$ is obtained from the exponential function definition $e^x = \mathop {\lim }\limits_{n \to \infty } {\left( {1 + x/n} \right)^n}$.
Hence, we can obtain the following equalities:
%\begin{equation}
${\rm{E}}\left[ {{Y_1}} \right] = {\rm{E}}\left[ {{Y_\infty}} \right]$ %\label{eq.sameE}
and $\ {f_{{Y_1}}}\left( y \right) = {f_{{Y_\infty}}}\left( y \right).$ %\label{eq.samePDF}
%\end{equation}
Since the transmit precoder at the BS is designed to match $\boldsymbol{h}_k$ independent with $\boldsymbol{f}_i^{\rm{H}}\boldsymbol{G}_i$,
the moments of instantaneous SNR at the uRUE will be a monotonic function with respect to $M$.
Therefore, $\gamma_i^{\rm{uR}}$ can be approximated as the chi-squared distributed random variable given in (\ref{eq.Chi}) regardless of $M$.
%because the SNR is a monotonic function in terms of $M$.
By using the characteristic of a chi-squared distribution, $\gamma_i^{\rm{uR}}$ has the following moment generating function (MGF) \cite{ref.Proakis2008digital}:
%\begin{equation}
$M_{\gamma_i^{\rm{uR}}} \left({\mu _p}, t \right) = {\frac{{{1}}}{{{1} -\frac{ N{N_{\rm{s}}}{E_{\rm{s}}}A{{\left( {{\mu _p}} \right)}^2t}}{N_0}}}}$. 
%\end{equation}
Based on this MGF, we can obtain the average SER of the proposed $M_0$-PSK signal at the uRUE as follows \cite{ref.Simon2005digital}:
\begin{equation}
P_{\rm{e}}=\frac{1}{{{2^{{M_o}}}}}\sum\limits_{p = 1}^{{2^{{M_o}}}} {\int_0^{\pi - \frac{\Delta_\mu}{2}} M_{\gamma_i^{\rm{uR}}} \left( {\mu _p},\frac{{ - {{\sin }^2}\left( {{\Delta_\mu / 2}} \right)}}{{{{\sin }^2}\theta }} \right) d\theta }.
\end{equation}

\vspace{-0.4cm}
\section*{Appendix C \\\vspace{-0.15cm} Proof of Proposition 2}\vspace{-0.1cm}
The integer solution $q_k^{t,f}$ always has a value of 0 or 1 for the entire range of $P^{t,f}$.
If we can calculate the optimal $P^{t,f}$ that maximizes $\bar R_k^{t,f}$ for each $q_k^{t,f} =1$, $k \in \cal K$ and $f \in \cal F$ (i.e., $q_i^{t,f} = 0$, $\forall i\ne k \in \cal K$ and $f \in \cal F$),
we can jointly find the optimal $P^{t,f}$ and $q_k^{t,f}$ by comparing those maximum $\bar R_k^{t,f}$ values.
Since $\bar \gamma_{{\rm{d}}k}^{t,f}$, $\bar \gamma_{{\rm{a}}i}^{t,f}$, and $\bar \gamma_{{\rm{r}}k}^{t,f}$ in (\ref{eq.newRtk}) are strictly concave functions with respect to $P^{t,f}$,
$\bar R_k^{t,f} - \mu P^{t,f}$ is also a concave function with respect to $P^{t,f}$.
For $k \in \cal R$, we can readily find the maximizer $\hat P_k^{t,f}$ by letting the first derivative be zero under consideration of the power constraint in (\ref{eq.Pcons}):
\begin{equation}
\hat P_k^{t,f} = \left[ {\frac{{1 + {\lambda _k}}}{\mu } - \frac{{{N_0}}}{{{{\Big| {{{\left( {{\bf{f}}_k^{t,f}} \right)}^{\rm{H}}}{\bf{\hat \Phi }}_k^{t,f}{\bf{\hat g}}_k^{t,f}} \Big|}^2}}}} \right]_0^{{P_{{\rm{max}}}}}, k \in {\cal R}, f  \in \cal F,\label{eq.Popt}\vspace{-0.1cm}
\end{equation}
where $\left[ x \right]_a^b = a$ if $x < a$, $\left[ x \right]_a^b = b$ if $x>b$, and otherwise, $\left[ x \right]_a^b = x$.
For $k \in \cal D$, we can also obtain the globally optimal $\hat P_k^{t,f}$ using a simple gradient method under the constraint in (\ref{eq.Pcons}).
From those results of $\hat P_k^{t,f}$, $\forall k \in \cal K$ and $\forall f \in \cal F$, we can find the optimal $\hat q_k^{t,f}$ as in (\ref{eq.optq}) and the corresponding $\hat P_k^{t,f}$ is the optimal transmission power, which completes the proof.
\vspace{-0.2cm}

\bibliographystyle{IEEEtran}
\bibliography{IEEEabrv,myBiB}

\end{document}